\documentclass[11pt,final]{article}

\usepackage[margin=2.5cm]{geometry}
\usepackage{amsfonts}
\usepackage{amssymb}
\usepackage{amsxtra}
\usepackage[english]{babel}
\usepackage{chemarr}
\usepackage{color}
\usepackage{epsfig}
\usepackage{graphicx}
\usepackage{latexsym}
\usepackage{natbib}
\usepackage{psfrag}
\usepackage{verbatim}
\usepackage{cancel}

\def\bone{{\bf 1}}
\def\bzero{{\bf 0}}

\def\Bkappa{\mbox{\boldmath$\kappa$}}

\def\Bsigma{\mbox{\boldmath$\sigma$}}

\def\Bvarphi{\mbox{\boldmath$\varphi$}}

\def\bC{\mbox{\boldmath$ C$}}
\def\bD{\mbox{\boldmath$ D$}}

\def\bF{\mbox{\boldmath$ F$}}

\def\bM{\mbox{\boldmath$ M$}}
\def\bN{\mbox{\boldmath$ N$}}

\def\bP{\mbox{\boldmath$ P$}}
\def\bQ{\mbox{\boldmath$ Q$}}

\def\bV{\mbox{\boldmath$ V$}}

\def\bX{\mbox{\boldmath$ X$}}

\def\ba{\mbox{\boldmath$ a$}}

\def\bc{\mbox{\boldmath$ c$}}

\def\be{\mbox{\boldmath$ e$}}

\def\bg{\mbox{\boldmath$ g$}}

\def\bm{\mbox{\boldmath$ m$}}
\def\bn{\mbox{\boldmath$ n$}}

\def\bq{\mbox{\boldmath$ q$}}

\def\bu{\mbox{\boldmath$ u$}}
\def\bv{\mbox{\boldmath$ v$}}

\def\bx{\mbox{\boldmath$ x$}}

\begin{document}

\title{The micromechanics of fluid-solid interactions during growth in
  porous soft biological tissue}
\author{H. Narayanan\footnote{Research Assistant, Department of
    Mechanical Engineering}\and E. M. Arruda\footnote{Professor,
    Department of Mechanical Engineering and Program in Macromolecular
    Science and Engineering} \and K. Grosh\footnote{Professor,
    Department of Mechanical Engineering and Department of Biomedical
    Engineering} \and K. Garikipati\footnote{Associate Professor,
    Department of Mechanical Engineering and Michigan Center for
    Theoretical Physics}} 

\date{University of Michigan, Ann Arbor}

\maketitle

\begin{abstract}
In this paper we address some modelling issues related to biological
growth. Our treatment is based on a recently-proposed, general
formulation for growth within the context of Mixture Theory (\emph{Journal
of the Mechanics and Physics of Solids}, \textbf{52}, 2004,
1595--1625). We aim to enhance this treatment by making it more
appropriate for the biophysics of growth in porous soft tissue, specifically
tendon. This involves several modifications to the mathematical
formulation to represent the reactions, transport and mechanics, and
their interactions. We also reformulate the governing differential
equations for reaction-transport to represent the incompressibility
constraint on the fluid phase of the tissue. This revision enables a
straightforward implementation of numerical stabilisation for the
hyperbolic, or ad\-vec\-tion-dom\-in\-ated, limit. A finite element
implementation employing an operator splitting scheme is used to solve
the coupled, non-linear partial differential equations that arise from
the theory.  Motivated by our experimental model, an {\em in vitro}
scaffold-free engineered tendon formed by self-assembly of tendon
fibroblasts (\emph{Tissue Engineering}, \textbf{10}, 2004, 755--761), we
solve several numerical examples 
demonstrating biophysical aspects of tissue growth, and the improved
numerical performance of the models.  
\end{abstract}

\section{Introduction}
\label{sec:1}

\emph{Growth} involves the addition or depletion of mass in biological
tissue. Growth occurs in combination with
\emph{remodelling}, which is a change in microstructure, and possibly
with \emph{morphogenesis}, which is a change in form in the embryonic
state. The physics of these processes are quite distinct, and for
modelling purposes can, and must, be separated. Our previous work
\citep{growthpaper}, upon which we now seek to build, drew in some
measure from \citet{CowinHegedus:76, EpsteinMaugin:2000}, and
\citet{TaberHumphrey:2001}, and was focused upon a comprehensive
account of the coupling between transport and mechanics. The origins
of this coupling were traced to the balance equations, kinematics and
constitutive relations. A major contribution of that work was the
identification and discussion of several driving forces for transport
that are thermodynamically-consistent, in the sense that specification
of these relations does not violate the Clausius-Duhem dissipation
inequality. 

There have been a number of significant papers on biological growth
and remodelling in the last 7--8 years of which we touch upon some,
whose approaches are either similar to ours in some respects or differ
in important ways. \citet{HumphreyRajagopal:02} provided a
mathematical treatment of \emph{adaptation} in a tissue, which includes 
growth and remodelling in the sense of this paper. The authors
identified adaptation as perhaps the most important mechanical
characteristic of biological tissue. They introduced the notion of evolving
natural configurations to model the state of material deposited at
different instants in time. The treatment of the growth part of the
deformation gradient in this paper bears some resemblance to this
idea, although a detailed development has not been pursued here. The
focus, instead, is on detailing some aspects of the problem that
derive from treatment of the tissue as a porous medium, or as a
mixture of interacting species. \citet{PreziosiFarina:2002} developed an
extension to the classical Darcy's Law to incorporate mass exchanges
between reacting species. This consideration is relevant to growth
problems; however, in our opinion, these issues were subsumed in
\citet{growthpaper}, upon which this paper is based. Many of the ideas
employed here are applicable to tumour growth problems; however, due
to our current focus on tendon, we do not include phenomena
such as angiogenesis and cell migration \citep[see for
  example][]{Brewardetal:2003}. The changes in concentration that
occur with growth tend to cause swelling or contraction of the
tissue. This phenomenon has been accounted for previously by us in fields unrelated to
Biology, using the idea of thermal expansion. See, for
example, \citet{Rao2:00} and \citet{Garikipatietal:01}, which, too,
are probably not the first instances of this idea. In the literature
on biological 
growth this connection was made by \citet{KlischHoger:2003}.

In the present paper, we seek to restrict the range of
physically-admissible models in order to gain greater
physiological relevance for modelling growth in soft biological
tissue. We also include one improvement in the mathematical/numerical treatment:
The advection-diffusion equations for mass transport
require numerical stabilisation in the advection-dominated regime
(the hyperbolic limit). We draw upon the enforcement of the
incompressibility limit for the fluid phase to facilitate this
development. Below, we briefly introduce each aspect that we have
considered, but postpone details until relevant sections in the paper.

\begin{itemize}
\item[\textbullet] For a tissue undergoing finite strain, the
  transport equations can be formulated, mathematically, in terms of
  concentrations with respect to either the reference or current
  (deformed) configuration. However, the physics of fluid-tissue
  interactions and the imposition of relevant boundary conditions is
  best understood and represented in the current configuration.

\item[\textbullet] The state of saturation is crucial in determining
  whether the tissue swells or shrinks with infusion/expulsion of
  fluid. This aspect has been introduced into the formulation.

\item[\textbullet] The fluid phase, whether slightly compressible or
  incompressible, can develop compressive stress without
  bound. However, it can develop at most a small tensile stress
  \citep{cavitationchris}, having implications for the stiffness of
  the tissue in tension as against compression. Although 
  this also has implications for void formation through cavitation,
  the ambient pressure in the tissue under normal physiological
  conditions ensures that this manifests itself only as a reduction in
  compressive pressure.

\item[\textbullet] When modelling transport, it is common to assume
  Fickean diffusion \citep{KuhlSteinmann:02}. This implies the
  existence of a mixing entropy due to the configurations available to
  molecules of the diffusing species at fixed values of the
  macroscopic concentration. The state of fluid saturation directly
  influences its mixing entropy.

\item[\textbullet] If fluid saturation is maintained, void formation
  in the pores is disallowed even under an increase in the pores'
  volume. This has implications for the fluid exchanges between a
  deforming tissue and a fluid bath with which it is in contact.

\item[\textbullet] Recognising the incompressibility of the fluid
  phase, it is common to treat soft biological tissue as either
  incompressible or nearly-incompressible \citep{Fung:1993}. At the
  scale of the pores (the microscopic scale in this case), however, a
  distinction exists in that the fluid is exactly (or nearly)
  incompressible, while the porous solid network is not obviously
  incompressible.

\item[\textbullet] In \citet{growthpaper}, the acceleration of the
  solid phase was included as a driving force in the constitutive
  relation for the flux of other phases. However, acceleration is not
  frame-invariant and its use in constitutive relations is
  inappropriate.

\item[\textbullet] Chemical solutes in the extra-cellular fluid are
  advected by the fluid velocity and additionally undergo transport
  under a chemical potential gradient relative to the fluid. In the
  hyperbolic limit, where advection dominates, spatial instabilities
  emerge in numerical solutions of these transport equations
  \citep{Brooks:82, Paper6}. Numerical stabilisation of the equations
  is intimately tied to the mathematical representation of fluid
  incompressibility.

\item[\textbullet] The modelling of solid-fluid mechanical coupling
  carries strong implications for the stiffness of tissue response,
  the nature of fluid transport, and since nutrients are dissolved in
  the fluid, ultimately for growth. We present upper and lower bounds
  for this problem and computations of coupled boundary value problems with
  these bounds.
\end{itemize}

These issues are treated in detail in relevant sections of the paper,
which is laid out as follows: Balance equations and kinematics are
discussed in Section~\ref{sec:2}, constitutive relations for
reactions, transport and mechanics in Section \ref{sec:3}, and
numerical examples are presented in Section
\ref{numericalimplementation}. Conclusions are drawn in
Section~\ref{sec:5}.

\section{Balance equations and kinematics of growth}
\label{sec:2}

In this section, the coupled, continuum balance equations governing
the behaviour of growing tissue are summarised and specialised as
outlined in Section~\ref{sec:1}. For detailed continuum mechanical
arguments underlying the equations, the interested reader is directed
to \citet{growthpaper}.

\begin{figure}[ht]
  \centering
  \psfrag{A}{\small$\bX$}
  \psfrag{F}{\small$\bx$}
  \psfrag{B}{\renewcommand{\baselinestretch}{1.5}\small$\Pi^\iota$}
  \psfrag{G}{\small$\pi^\iota$}
  \psfrag{E}{\small$\Bvarphi$}
  \psfrag{C}{\small$\Omega_0$}
  \psfrag{H}{\small$\Omega_t$}
  \psfrag{D}{\small$\bN\cdot\bM^\iota$}
  \psfrag{I}{\small$\bn\cdot\bm^\iota$}
	 {\includegraphics[width=7.50cm]{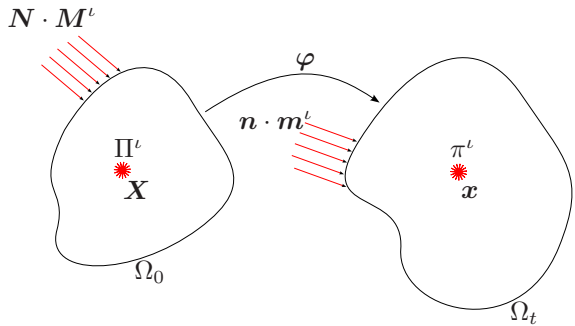}}
	 \caption{The tissue as a continuous medium with growing and
           diffusing species.}
	 \label{cp}
\end{figure}

The tissue of interest is an open subset of $\mathbb{R}^3$ with a
piecewise smooth boundary. At a reference placement of the tissue,
$\Omega_0$, points in the tissue are identified by their reference
positions, $\bX \in \Omega_0$. The motion of the tissue is a
sufficiently smooth bijective map $\Bvarphi: \overline{\Omega}_0
\times [0,T] \rightarrow \mathbb{R}^3$, where $\overline{\Omega}_0 :=
\Omega_0 \cup \partial\Omega_0$; $\partial\Omega_0$ being the boundary
of $\Omega_0$. At a typical time $t \in [0,T]$, $\Bvarphi(\bX,t)$ maps
a point $\bX$ to its current position, $\bx$. In its current
configuration, the tissue occupies a region $\Omega_t = \Bvarphi_t
(\Omega_0)$. These details are depicted in Figure~\ref{cp}. The
deformation gradient $\bF := \partial \Bvarphi/ \partial\bX$ is the
tangent map of $\Bvarphi$.

The tissue consists of numerous species, of which the following
 groupings are of importance for the models: A solid species,
 consisting of solid \emph{collagen fibrils} and \emph{cells},\footnote{At this
 point, we do not distinguish the solid species further. This is a
 good approximation to the physiological setting for tendons, which
 are relatively acellular and whose dry mass consists of up to 75\%
 collagen \citep{Nordinetal:2001}.} denoted by $\mathrm{c}$, an
 extra-cellular \emph{fluid} species denoted by $\mathrm{f}$ and
 consisting primarily of water, and \emph{solute} species, consisting
 of precursors to reactions, byproducts, nutrients, and other
 regulatory chemicals. A generic solute will be denoted by
 $\mathrm{s}$. In what follows, an arbitrary species will be denoted
 by $\iota$, where $\iota = \mathrm{c,f,s}$.

The fundamental quantities of interest are mass concentrations,
$\rho_0^\iota(\bX,t)$. These are the masses of each species per unit
system volume in $\Omega_0$. Formally, these quantities can also be
thought of in terms of the maps $\rho_0^\iota: \overline{\Omega}_0
\times [0,T] \rightarrow \mathbb{R}$, upon which the formulation
imposes some smoothness requirements. By definition, the total {\em
material density} of the tissue at a point is a sum of these
concentrations over all species $\sum\limits_{\iota}\rho_0^\iota =
\rho_0$. Other than the solid species, $\mathrm{c}$, all phases have
mass fluxes, $\bM^\iota$.\footnote{Currently, we do not consider
  certain physiological 
processes, such as the migration of fibroblasts within the
extra-cellular matrix during wound healing, which may otherwise be
modelled as mass transport.} These are mass flow rates per unit
cross-sectional area in the reference configuration \emph{defined
relative to the solid phase}. The species have mass sources (or
sinks), $\Pi^\iota$.

\subsection{Balance of mass for an open system}
\label{bomass} As a result of mass transport (via the flux terms) and
inter-conversion of species (via the source/sink terms) introduced
above, the concentrations, $\rho_0^\iota$, change with time. In
local form, the balance of mass for an arbitrary species in the
reference configuration is

\begin{equation}
\frac{\partial\rho_0^\iota}{\partial t} = \Pi^\iota -
\mathrm{\small{DIV}}[\bM^\iota],\;\forall\,\iota,
\label{massbalance1}
\end{equation}

\noindent recalling that, in particular, $\bM^{c} = \bzero$. Here,
$\mathrm{\small{DIV}[\bullet]}$ is the divergence 
operator in the reference configuration. The functional forms of
$\Pi^\iota$ are abstractions of the underlying biochemistry,
physiologically relevant examples of which are discussed in
Section~\ref{sources}, and the fluxes, $\bM^\iota$, are determined
from the thermodynamically-motivated constitutive relations described
in Section~\ref{flux const}.

The behaviour of the entire system can be determined by summing
\mbox{Equation~(\ref{massbalance1})} over all species $\iota$.
Additionally, sources and sinks satisfy the relation

\begin{equation}
\sum\limits_\iota\Pi^\iota = 0, \label{sourcebalance}
\end{equation}

\noindent which is consistent \citep{growthpaper} with the Law of Mass
Action for reaction rates and with Mixture Theory
\citep{TruesdellNoll:65}.

\subsubsection{The role of mass balance in the current
configuration}\label{curr-ref-mb}

In order to proceed, we must first introduce the central kinematic
assumption underlying the 
formulation: We assume that the pore structure deforms with the
collagenous phase. Therefore, the deformation gradient, $\bF$, is
common to c and the fluid-filled pore spaces. Furthermore, in what
follows, we will treat the fluid as ideal and nearly-incompressible,
i.e. as elastic (Section \ref{compfluid}). This combination of
kinematic and constitutive assumptions to be elaborated upon, implies
that the stress in the fluid phase is determined by the elastic part of
$\bF$ (see Sections \ref{growthkinem} and
\ref{compfluid}). For clarity we denote it as
$\bF^{\mathrm{e}^\mathrm{f}}$. Importantly, the pore-filling fluid under stress
can also undergo transport relative to the pore network; i.e.,
relative to the collagenous phase. This is the fluid flux, denoted by
$\bM^\mathrm{f}$ in the reference configuration. Note that the
specification of constitutive relations for the flux is still open at
this point in the discussion. At the outset, we preclude stress in any
of the solute species, s. Only the solid collagen and fluid bear
stress. 

\begin{figure}
\centering
\includegraphics[width=7.50cm]{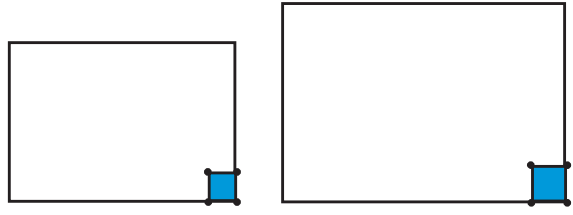}
\caption{If the pore structure at the boundary deforms with the
tissue and this boundary is in contact with a fluid bath, the
fluid concentration with respect to the current configuration,
i.e., $\rho^\mathrm{f}$, remains constant.}\label{fbc}
\end{figure}

Although the initial/boundary value problem of mass transport can be
consistently posed in the reference configuration, the evolving
current configuration, $\Omega_t$, is of greater interest from a
physical standpoint for growth problems. It follows from the
discussion in the preceding paragraph that the shape and size of
pores in $\Omega_t$ is determined by $\bF$. Therefore, at
the boundary, the fluid concentration with 
respect to $\Omega_t$ remains constant if the boundary is in contact
with a fluid bath.  Accordingly, this is the appropriate Dirichlet
boundary condition to impose under normal physiological
conditions. This is shown in an idealised manner in Figure~\ref{fbc}.

In the interest of applying boundary conditions (either specification
of species flux or concentration) that are physically meaningful, we
use the local form of the balance of mass in the current
configuration,

\begin{equation}
\frac{\mathrm{d}\rho^\iota}{\mathrm{d}t} = \pi^\iota-
\mathrm{\small{div}}[\bm^\iota] - \rho^\iota
\mathrm{\small{div}}[\bv],\;\forall\,\iota, \label{massbalcurr}
\end{equation}

\noindent where $\rho^\iota(\bx,t),\pi^\iota(\bx,t)$, and
$\bm^\iota(\bx,t)$ are the current mass concentration, source and mass
flux of species $\iota$ respectively and $\bv(\bx,t)$ is the velocity
of the solid phase. They are related to corresponding reference
quantities as $\rho^\iota = \left(\mathrm{det} \left(\bF\right)
\right)^{-1} \rho_0^\iota$, $\pi^\iota = \left(\mathrm{det}
\left(\bF\right) \right)^{-1} \Pi^\iota$ and $\bm^\iota =
\left(\mathrm{det} \left(\bF\right) \right)^{-1} \bF \bM^\iota$. The
spatial divergence operator is $\mbox{\small{div} [\textbullet]}$, and
the left hand-side in Equation~(\ref{massbalcurr}) is the material
time derivative relative to the solid, which may be written explicitly as
$\frac{\partial}{\partial t}\vert_X$, implying that the reference
position of the solid collagenous skeleton is held fixed.

\subsection{The kinematics of growth induced by changes in concentration}
\label{growthkinem}

\begin{figure}[ht]
  \centering
  \psfrag{A}{\small $\Omega_0$}
  \psfrag{B}{\small $\Omega^\ast$}
  \psfrag{C}{\small $\Omega_t$}
  \psfrag{D}{\small $\bX$}
  \psfrag{E}{\small $\bX^\ast$}
  \psfrag{F}{\small $\bx$}
  \psfrag{G}{\small $\bF^{\mathrm{g}^{\mathrm{\iota}}}$}
  \psfrag{H}{\small $\widetilde{\bF}^{\mathrm{e}^{\mathrm{\iota}}}$}
  \psfrag{I}{\small $\widetilde{\bF}$}
  \psfrag{J}{\small $\overline{\bF}^\mathrm{e}$}
  \psfrag{K}{\small $\bF$}
  \psfrag{L}{\small $\Bkappa$}
  \psfrag{M}{\small $\bu^\ast$}
  \psfrag{N}{\small $\Bvarphi$}
         {\includegraphics[width=7.50cm]{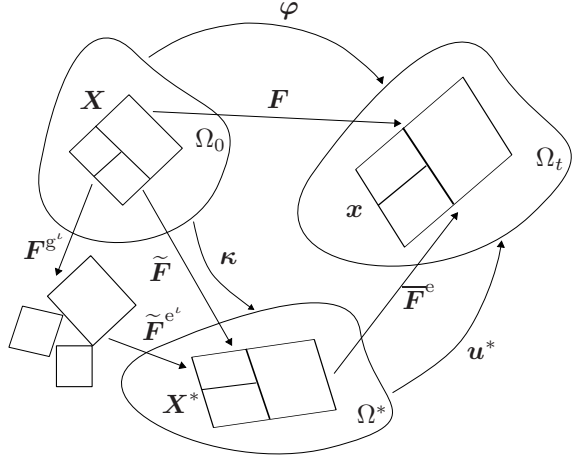}}
     \caption{The kinematics of growth, which holds for $\iota =
     \mathrm{c,f}$.} 
     \label{growthkine}
\end{figure}

Local volumetric changes are associated with changes in the
concentrations of the solid collagen and fluid, $\iota =
     \mathrm{c,f}$. If the material of the solid collagen or fluid remains
stress free, it swells with an
increase in concentration (mass of the species per unit system
volume), and shrinks as its concentration decreases. This leads to the
notion of a \emph{growth deformation gradient}. One aspect of the
coupling between mass transport and mechanics stems from this
phenomenon. In the setting of finite strain kinematics, the total
deformation gradient, $\bF$, is decomposed into the growth component
of the solid collagen, $\bF^{\mathrm{g}^\mathrm{c}}$, a
\emph{geometrically-necessitated elastic component} 
accompanying growth, $\widetilde{\bF}^{\mathrm{e}^\mathrm{c}}$ and an \emph{additional elastic component due
to external stress}, $\overline{\bF}^{\mathrm{e}^\mathrm{c}}$. Later,
we will write $\bF^{\mathrm{e}^\mathrm{c}} =
\overline{\bF}^{\mathrm{e}^\mathrm{c}}\widetilde{\bF}^{\mathrm{e}^\mathrm{c}}$.
This split is analogous to the classical
decomposition of multiplicative plasticity \citep{Lee:69} and is
similar to the approach followed in existing literature on biological
growth \citep[see for
  e.g.][]{Klischetal:2001,TaberHumphrey:2001,AmbrosiMollica:2002}. As
explained in Section \ref{curr-ref-mb}, we assume that the
fluid-filled pores also deform with $\bF$, and that a component,
$\bF^{\mathrm{e}^\mathrm{f}}$, of this total deformation gradient
tensor, determines the fluid stress. We also assume a fluid growth
component, $\bF^{\mathrm{g}^\mathrm{f}}$, which we elaborate below,
and that $\bF^{\mathrm{e}^\mathrm{f}}\bF^{\mathrm{g}^\mathrm{f}} =
\bF$. As with the solid collagen we admit $\bF^{\mathrm{e}^\mathrm{f}}
=
\overline{\bF}^{\mathrm{e}^\mathrm{f}}\widetilde{\bF}^{\mathrm{e}^\mathrm{f}}$,
the sub-components carrying the same interpretation as for the solid
collagen. However, we do not explicitly use this last decomposition.

The elastic-growth decomposition is visualised in \mbox{Figure~\ref{growthkine}}.
Assuming that the volume changes associated with growth described
above are isotropic, a simple form for the growth part of the
deformation gradient tensor is

\begin{equation}
\bF^{\mathrm{g}^\iota} = \left(
  \frac{\rho_0^\iota}{\rho_{0_{\mathrm{ini}}}^\iota} \right)^
  {\frac{1}{3}} 
{\bf 1},\quad \iota = \mathrm{c,f}
\label{isotropicgrowth} 
\end{equation} 

\noindent where
$\rho_{0_{\mathrm{ini}}}^\iota(\bX)$ is the reference concentration at the
initial time, and {\bf 1} is the second-order isotropic
tensor.\footnote{This choice is only the simplest possible. Given the
  highly directional micro-structure and mechanical properties of many
  tissues, it seems likely that anisotropic growth is actually 
  more common. Wolff's Law for bone growth is one example. This is a
  topic of ongoing investigation, and one that we will report on in
  greater detail in a future communication.} In 
the state, $\bF = \bF^{\mathrm{g}^\iota}$, the species would be stress
free. The kinematics being local, the
action of $\bF^{\mathrm{g}^\iota}$ alone can result in
incompatibility, which is eliminated by the geometrically-necessary
elastic deformation
$\widetilde{\bF}^{\mathrm{e}^{\mathrm{\iota}}}$, which causes an
internal, self-equilibrated stress. The component
$\overline{\bF}^{\mathrm{e}^\iota}$ is associated with the external stress.

\subsubsection{Saturation and tissue swelling}\label{satswel}

\begin{figure}[ht]
\centering 
  \psfrag{A}{\small A}
  \psfrag{B}{\small B}
  \psfrag{C}{\small B}
  \psfrag{D}{\small C}
  \psfrag{E}{\small Unsaturated}
  \psfrag{F}{\small Saturated}
{\includegraphics[width=7.50cm]{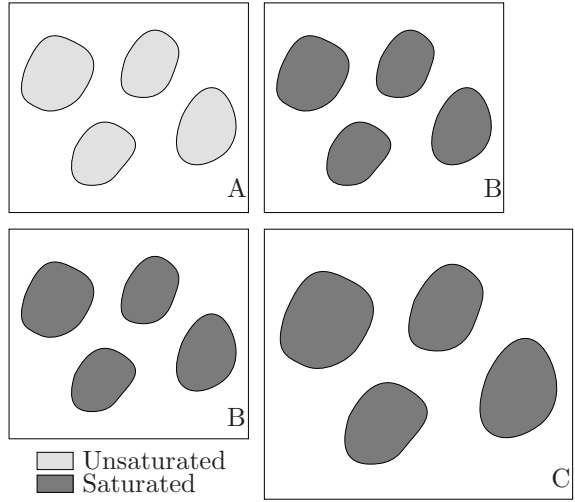}}

\caption{Unsaturated tissue in the current configuration (A) allows
  influx of fluid 
  without swelling until it is completely saturated (B). Initially
  saturated tissue (B), in general, swells with influx of fluid (C).}

\label{satswelfig}
\end{figure}

\noindent The degree of saturation of the solid phase plays a
fundamental role in determining whether the tissue responds to an
infusion (expulsion) of fluid by swelling (shrinking). In 
particular, the isotropic swelling law defined by Equation
(\ref{isotropicgrowth}) has to be generalised to treat the case in
which the solid phase is not saturated by fluid.

Figure~\ref{satswelfig} schematically depicts two possible
scenarios. If the tissue is unsaturated in its current configuration,
as in A, then, on a microscopic scale, it 
contains unfilled voids. It is thus capable of allowing an influx of
fluid, which tends to increase its degree of saturation until fully
saturated, as in  B. This increase 
does not cause swelling of the tissue in the local stress-free state, as
there is free volume for 
incoming fluid to occupy. However, once the tissue is
saturated in the current configuration, an increase in the
fluid content causes swelling in the stress-free state, as depicted in
C, since there is no free volume for the 
entering fluid to occupy. It is this second case that is modelled by
(\ref{isotropicgrowth}). It is worth emphasizing that this argument
holds for $\bF^{\mathrm{g}^\mathrm{f}}$, which is the local
stress-free state of deformation of the fluid-containing pores at a point. The
actual deformation gradient, $\bF =
\bF^{\mathrm{e}^\mathrm{f}}\bF^{\mathrm{g}^\mathrm{f}}$, also depends
on the the elastic part, $\bF^{\mathrm{e}^\mathrm{f}}$, which is
determined by the constitutive response of the fluid. Under stress, an
incompressible fluid will have
$\mathrm{det}\bF^{\mathrm{e}^\mathrm{f}} = 1$ and therefore a
fluid-saturated tissue will swell with fluid influx, $\mathrm{det}\bF
= \mathrm{det}\bF^{\mathrm{g}^\mathrm{f}} > 1$. A compressible fluid
may have $\mathrm{det}\bF^{\mathrm{e}^\mathrm{f}} < 1$ allowing
$\mathrm{det}\bF < 1$ even with
$\mathrm{det}\bF^{\mathrm{g}^\mathrm{f}} >1$. Even in this case,
however, in the stress-free state there will be swelling.

Therefore, for the fluid phase, the isotropic swelling law can be
extended to the unsaturated case by 
introducing a degree of saturation, $\tilde{v}^\iota$, defined in the
current configuration, $\Omega_t$. We have $\tilde{v}^\iota =
\rho^\iota/\tilde{\rho}^\iota$, where 
$\tilde{\rho}^\iota$ is the intrinsic density in $\Omega_t$ and is
given by $\tilde{\rho}^\iota =
\tilde{\rho}^\iota_0/\mathrm{det}\bF$. Note that the intrinsic
reference density, $\tilde{\rho}^\iota_0$, is a material property. Upon
solution of the 
mass balance equation (\ref{massbalcurr}) for $\rho^\iota$, the
species volume fractions, $\tilde{v}^\iota$, can therefore be computed
in a straightforward fashion. The sum of these
volume fractions is our required measure of saturation defined in
$\Omega_t$. Also, recognizing
that for the dilute solutions obtained with
physiologically-relevant solute concentrations, the saturation
condition is very well approximated by $\tilde{v}^\mathrm{f} +
\tilde{v}^\mathrm{s} = 1$, we proceed to
redefine the fluid growth-induced component of the pore deformation gradient
tensor as follows:

\begin{equation}
\bF^{\mathrm{g}^\mathrm{f}} = \left\{ \begin{array}{ll}  \left(
\frac{\rho_0^\mathrm{f}}{\rho_{0_{\mathrm{sat}}}^\mathrm{f}} \right)^
{\frac{1}{3}}{\bf 1},&
\tilde{v}^\mathrm{f} + \tilde{v}^\mathrm{s} = 1 \\ {\bf 1},& \mathrm{otherwise.}
\end{array}\right.
\label{saturation}
\end{equation}

\noindent In (\ref{saturation}) $\rho_{0_{\mathrm{sat}}}^\mathrm{f}$
is the reference concentration value at which the tissue attains saturation
in the current configuration.

With this redefinition of $\bF^{\mathrm{g}^\mathrm{f}}$ it is implicit
that $\tilde{v}^\mathrm{f} + \tilde{v}^\mathrm{s} > 1$ is
non-physical. Saturation holds in the sense that $\tilde{v}^\mathrm{f} +
\tilde{v}^\mathrm{s} = 1$, and it actually allows $\sum_\iota
\tilde{v}^\iota > 1$ if the sum is over all species. It has been
common in the soft tissue literature to assume that, 
under normal physiological 
conditions, soft tissues are fully saturated by the fluid and
\mbox{Equation~(\ref{isotropicgrowth})} is appropriate for $\iota =
\mathrm{f}$. However, this treatment of saturation and swelling
induced by the fluid phase is necessary background for Section
\ref{tensionfluid} where we 
discuss the response of the fluid phase under tension. This treatment
also holds relevance for partial drying,
which \emph{ex vivo} or \emph{in vitro} tissue may be subject to under
certain laboratory conditions, and is central to the mechanics of
drained porous media other than biological tissue, most prominently,
soils. 

\subsection{Balance of momenta}
\label{bomom}

In soft tissues, the species production rate and flux that appear on
the right hand-side in Equations~(\ref{massbalance1}) and
(\ref{massbalcurr}), are strongly 
dependent on the local state of stress. To correctly model this
coupling, the balance of linear momentum should be solved to determine
the local state of strain and stress.

The deformation of the tissue is characterised by the map
$\Bvarphi(\bX,t)$. Recognising that, in tendons, the solid collagen
fibrils and fibroblasts do not undergo mass
transport, the material velocity of this species, $\bV =
\partial\Bvarphi/\partial t$, is used as the primitive variable for
mechanics. Each remaining species can undergo mass transport relative
to the solid collagen. For this purpose, it is useful to define the
material velocity of a 
species $\iota$ \emph{relative to the solid skeleton} as: $\bV^\iota =
(1/\rho_0^\iota)\bF\bM^\iota$. Thus, the total material velocity of a
species $\iota$ is $\bV+\bV^\iota$.

The total first Piola-Kirchhoff stress tensor, $\bP$, is the sum of
the partial stresses $\bP^{\,\iota}$ (borne by a species $\iota$) over all
the species present. Recognizing that solutes in low concentrations,
and do not bear 
  appreciable stress, the partial stresses and momentum balance
  equation are defined only for the solid collagen and fluid
  phases. With the introduction 
  of these quantities, the 
balance of linear momentum in local form over $\Omega_0$ for solid
collagen and fluid is,

\begin{equation}
\begin{split}
\rho_0^\iota\frac{\partial}{\partial t}\left(\bV+\bV^\iota\right) &
=\rho^\iota_0\left(\bg+\bq^\iota\right) +
\mathrm{\small{DIV}}[\bP^{\,\iota}]\\ 
& \quad -\left(\mathrm{\small{GRAD}}\left[\bV+
  \bV^\iota\right]\right)\bM^\iota, \quad \iota = \mathrm{c,f}
\label{linearmombalance}
\end{split}
\end{equation} 

\noindent where $\bg$ is the body force per unit mass, and $\bq^\iota$
is an interaction term denoting the force per unit mass exerted upon
$\iota$ by all other species present. The final term with the
(reference) gradient denotes the contribution of the flux to the
balance of momentum. In practise, the relative magnitude of the fluid
mobility (and hence flux) is small, so the final term on the right
hand side of Equation~(\ref{linearmombalance}) is negligible,
resulting in a more classical form of the balance of
momentum. Furthermore, in the absence of significant acceleration of
the tissue during growth, the left hand-side can also be neglected,
reducing (\ref{linearmombalance}) to the quasi-static balance of
linear momentum.

The balance of momentum of the entire tissue is obtained by summing
Equation~(\ref{linearmombalance}) over $\iota = \mathrm{c,f}$. Additionally,
recognising that the rate of change of momentum of the entire tissue
is affected only by external agents and is independent of internal
interactions, the following relation arises.

\begin{equation}
\sum\limits_{\iota =
  \mathrm{c}}^\mathrm{f}\left(\rho^\iota_0\bq^\iota+\Pi^\iota
\bV^\iota 
\right)= 0. \label{qrelation}
\end{equation}

\noindent This is also consistent with Classical Mixture Theory
\citep{TruesdellNoll:65}. See \citet{growthpaper} for further
details on balance of linear momentum, and the formulation of
balance of angular momentum. We only note here that the latter
principle leads to a symmetric partial Cauchy stress,
$\Bsigma^\iota$ for each species in contrast with the unsymmetric
Cauchy stress of \cite{EpsteinMaugin:2000}.

\section{Constitutive framework and specific models}
\label{sec:3}

As is customary in field theories of continuum physics, the
Clausius-Duhem inequality is obtained by multiplying the Entropy
Inequality (the Second Law of Thermodynamics) by the temperature
field, $\theta$, and subtracting it from the Balance of Energy (the
First Law of Thermodynamics). We assume the Helmholtz free energy per
unit mass of species $\iota$ to have the form:\footnote{Conceivably,
  the mass-specific Helmholtz free energy of one species could be a function of the
  concentration of other species. Ion concentrations, for instance,
  can determine the state of osmotic tension of certain soft
  tissues. Therefore, this choice represents a 
  constitutive restriction.}
$\psi^{\iota} = \hat{\psi}^\iota(\bF^{\mathrm{e}^\iota},
\theta, \rho_0^\iota)$.  Substituting this in the Clausius-Duhem
inequality results in a form of this inequality that the specified
constitutive relations \emph{must not} violate. Only the valid
constitutive laws relevant to the examples that follow are listed
here. For details, see \cite{growthpaper}.

\subsection{An anisotropic network model based on entropic
elasticity}\label{wlcm}

The partial first Piola-Kirchhoff stress of collagen, modelled as a
hyperelastic material, is $\bP^{\,\mathrm{c}} = \rho_0^\mathrm{c} \partial
\psi^\mathrm{c}/ \partial\bF^{\mathrm{e}^\mathrm{c}}$. Recall that
$\bF^{\mathrm{e}^\mathrm{c}} = \bF\bF^{\mathrm{g}^{\mathrm{c}^{-1}}}$
is the elastic part, and
$\bF^{\mathrm{g}^\mathrm{c}}$ 
is the growth part, respectively, of the deformation gradient, of
collagen. Following Equation 
(\ref{isotropicgrowth}), if we were considering unidirectional growth
of collagen along a unit vector $\be$, we would have
$\bF^{\mathrm{g}^\mathrm{c}} = \frac{\rho^\mathrm{c}_{0}}
{\rho^\mathrm{c}_{0_{\mathrm{ini}}}} \be \otimes \be$, with
$\rho^\mathrm{c}_{0_{\mathrm{ini}}}$ denoting the initial
concentration of collagen at the point.

The mechanical response of tendons in tension is determined primarily
by their dominant structural component: highly oriented fibrils of
collagen. In our preliminary formulation, the strain energy density
for collagen has been obtained from hierarchical multi-scale
considerations based upon an entropic elasticity-based worm-like chain
(WLC) model \citep{KratkyPorod:49}. The WLC model has been widely used
for long chain single molecules, most prominently for DNA
\citep{MarkoSiggia:95,Riefetal:97,Bustamanteetal:2003}, and recently
for the collagen mono\-mer \citep{Sunetal:2002}. The central
parameters of this model are the chain's contour length, $L$, and
persistence length, $A$. The latter is a measure of its stiffness and
given by $A = \chi/k\theta$, where $\chi$ is the bending rigidity, $k$
is Boltzmann's constant and $\theta$ is the temperature. See
\citet{LandLif} for general formulation of statistical mechanics
models of long chain molecules. 

\begin{figure}
\psfrag{r}{\small $r$}
\psfrag{A}{\small $A$}
\psfrag{a}{\small $a$}
\psfrag{b}{\small $b$}
\psfrag{c}{\small $c$}
\psfrag{n}{\small$\bN_1$}
\psfrag{o}{\small$\bN_2$}
\psfrag{p}{\small$\bN_3$}
\centering {\includegraphics[width=7.50cm]{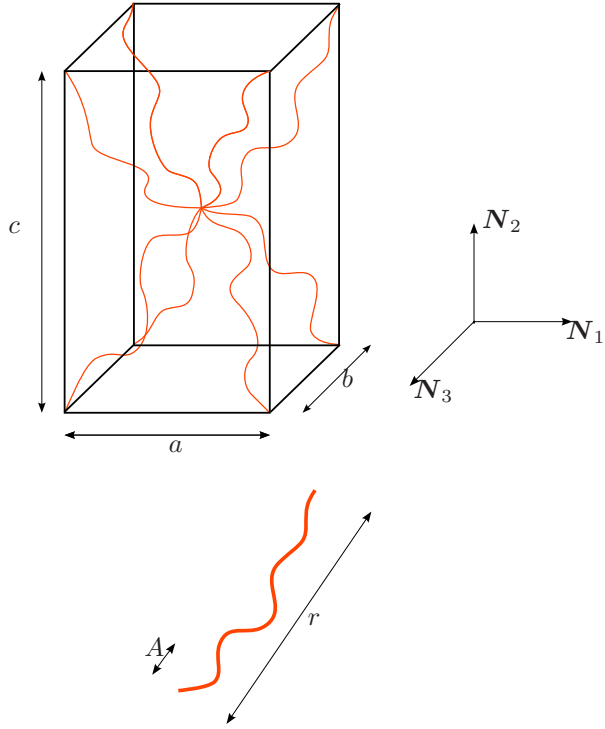}}
\caption{\small The eight-chain model incorporating worm-like
  chains.}\label{eightchain}
\end{figure}

To model a collagen network structure, the
WLC model has been embedded as a single constituent chain of an
eight-chain model \citep{Bischoffetal:2002, Bischoffetal1:2002},
depicted in \mbox{Figure~\ref{eightchain}}.  Homogenisation via
averaging then leads to a continuum Helmholtz free energy function,
$\hat{\psi}^\mathrm{c}$:\footnote{Under the isothermal conditions
  assumed here, $\hat{\psi}^\mathrm{c}$ is independent of
  $\theta$ in the 
  strain energy. Accordingly, we have the parametrisation
  ${\psi}^\mathrm{c}=\hat{\psi}^\mathrm{c}
  (\bF^{\mathrm{e}^\mathrm{c}},\rho^\mathrm{c}_{0})$ .}

\begin{equation}
\begin{split}
\rho^\mathrm{c}_{0}\hat{\psi}^\mathrm{c} (\bF^{\mathrm{e}^\mathrm{c}},\rho^\mathrm{c}_{0})
&= \frac{N k \theta}{4 A}\left(\frac{r^2}{2L} + \frac{L}{4(1-r/L)} -
\frac{r}{4}\right)\\ & +
\frac{\gamma}{\beta}({J^{\mathrm{e}^{\mathrm{c}^{-2\beta}}}} -1) +
\gamma{\bf 1}\colon(\bC^{\mathrm{e}^{\mathrm{c}}}-{\bf 1})\\ &-\frac{N
k \theta}{4\sqrt{2L/A}}\left(\sqrt{\frac{2A}{L}} + \frac{1}{4(1 -
\sqrt{2A/L})} -\frac{1}{4} \right) Z,\\ Z &=
\log\left(\lambda_1^{{\mathrm{e}}^{a^\mathrm{2}}}
\lambda_2^{{\mathrm{e}}^{b^\mathrm{2}}}
\lambda_3^{{\mathrm{e}}^{c^\mathrm{2}}}\right).
\label{wlcmeq}
\end{split}
\end{equation}

Here, $N$ is
the density of chains, and $a,b$ and $c$ are lengths of the unit cell
sides aligned with the principal stretch directions. The material
model is isotropic only if $a=b=c$.

The elastic stretches along the unit cell axes are, respectively,
denoted by
$\lambda_1^{\mathrm{e}},
\lambda_2^{\mathrm{e}}$ 
and 
$\lambda_3^{\mathrm{e}}$, $\bC^{\mathrm{e}^{\mathrm{c}}} =
\bF^{\mathrm{e}^{\mathrm{c}^{\mathrm{T}}}}
\bF^{\mathrm{e}^{\mathrm{c}}}$ is the elastic right Cauchy-Green
tensor of collagen. The factors $\gamma$ and $\beta$ control
the bulk compressibility of the model. The end to end chain length is
given by $r = \frac{1}{2}\sqrt{a^2\lambda_1^{\mathrm{e}^2} +
  b^2\lambda_2^{\mathrm{e}^2}+c^2\lambda_3^{\mathrm{e}^2}}$, where
$\lambda^\mathrm{e}_I = 
\sqrt{\bN_I\cdot\bC^{\mathrm{e}^{\mathrm{c}}}\bN_I}$, and $\bN_I,\,I =
1,2,3$ are the unit vectors along the three unit cell axes,
respectively. In our numerical simulations that appear below in
Section~\ref{numericalimplementation}, the numerical values used for
the parameters introduced in (\ref{wlcmeq}) are based on those in
\citet{kuhlremod05}.

\subsection{A nearly incompressible ideal fluid}
\label{compfluid}

In this preliminary work, the fluid phase is treated as nearly
incompressible and ideal, i.e., inviscid. The partial Cauchy
stress in the fluid is
\begin{equation}
\Bsigma^\mathrm{f} =
\mathrm{det}(\bF^{\mathrm{e}^\mathrm{f}})^{-1}
\bP^{\,\mathrm{f}}\bF^{\mathrm{e}^\mathrm{fT}} 
= h(\rho^\mathrm{f}){\bf 1},\label{Pf}
\end{equation}

\noindent where a large value of $h^\prime(\rho^\mathrm{f})$ ensures
near-in\-comp\-ress\-i\-bil\-i\-ty. 
 
\subsubsection{Response of the fluid in tension;
cavitation}\label{tensionfluid}

The response of the ideal fluid, as defined by \mbox{Equation
  (\ref{Pf})}, does not distinguish between tension and compression,
  i.e., whether 
$\mathrm{det}(\bF^{\mathrm{e}^\mathrm{f}}) \gtreqless 1$. Being
(nearly) incompressible, the fluid can develop compressive hydrostatic
  stress
without bound---a case that is modelled accurately. However, the fluid
can develop at most a small tensile hydrostatic stress
\citep{cavitationchris},\footnote{Where, we are referring to the fluid
  being subject to net tension, not a reduction in fluid compressive
  stress from reference ambient pressure.} and the tensile
stiffness is mainly from the collagen phase. This is not accurately
represented by (\ref{Pf}), which models a symmetric response in
tension and compression.

Here, we preclude all tensile load carrying by the fluid by limiting
$\mathrm{det}(\bF^{\mathrm{e}^\mathrm{f}}) \leq 1$. We first introduce
an additional component to the relation between 
deformation of the pore space, given by $\bF$, the fluid
stress-determining tensor, 
$\bF^{\mathrm{e}^\mathrm{f}}$ and the growth tensor for the fluid,
$\bF^{\mathrm{g}^\mathrm{f}}$. Consider the cavitation (void forming) tensor,
$\bF^{\mathrm{v}}$, defined by
  
\begin{equation}
 \bF^{\mathrm{e}^\mathrm{f}}\bF^{\mathrm{g}^\mathrm{f}}
 \bF^{\mathrm{v}} = \bF.
\label{fvoid}
\end{equation}

We restrict the formulation to include only saturated current
configurations at $t = 0$. Following Section \ref{satswel} we have
$\tilde{v}^\mathrm{f} + \tilde{v}^\mathrm{c} = 1$ at $t = 0$, the saturation
condition in $\Omega_t$ when solutes are at low
concentrations. At times $t > 0$ Equation (\ref{saturation}) holds for
$\bF^{\mathrm{g}^\mathrm{f}}$. If
$\mathrm{det}[\bF(\bF^{\mathrm{g}^\mathrm{f}})^{-1}] \le 1$ we set
$\bF^{\mathrm{e}^\mathrm{f}} = \bF(\bF^{\mathrm{g}^\mathrm{f}})^{-1}$
and $\bF^{\mathrm{v}} = {\bf 1}$ for no cavitation. Otherwise, since
$\mathrm{det}[\bF(\bF^{\mathrm{g}^\mathrm{f}})^{-1}] > 1$, we specify
$\bF^{\mathrm{e}^\mathrm{f}} =
\mathrm{det}[\bF(\bF^{\mathrm{g}^\mathrm{f}})^{-1}]^{-1/3}\bF(\bF^{\mathrm{g}^\mathrm{f}})^{-1}$
thus restricting
$\bF^{\mathrm{e}^\mathrm{f}}$ to be unimodular and allow cavitation by
writing $\bF^{\mathrm{v}} = \bF(\bF^{\mathrm{e}^\mathrm{f}}\bF^{\mathrm{g}^\mathrm{f}})^{-1}$.
These conditional relations are summarized as 

\begin{equation}
\bF^{\mathrm{e}^\mathrm{f}} = \left\{ \begin{array}{ll}
  \bF(\bF^{\mathrm{g}^\mathrm{f}})^{-1},\; \bF^{\mathrm{v}} = {\bf 1},&
 \mathrm{det}[\bF(\bF^{\mathrm{g}^\mathrm{f}})^{-1}] \le 1\\
  \mathrm{det}[\bF(\bF^{\mathrm{g}^\mathrm{f}})^{-1}]^{-1/3}\bF(\bF^{\mathrm{g}^\mathrm{f}})^{-1},
  & \\
  \qquad\qquad \bF^{\mathrm{v}} = \bF(\bF^{\mathrm{e}^\mathrm{f}}\bF^{\mathrm{g}^\mathrm{f}})^{-1}
  & \mathrm{otherwise.}
\end{array}\right.
\label{cavitation}
\end{equation}
 
\subsection{Constitutive relations for fluxes}
\label{flux const}

From \citet{growthpaper}, the constitutive relation for the flux of 
extra-cellular fluid relative to collagen in the reference
configuration takes the following form,

\begin{equation}
\bM^\mathrm{f} = \bD^\mathrm{f}\left(\rho_0^\mathrm{f}\bF^T\bg +
      \bF^T\mathrm{\small{DIV}}\left[\bP^{\,\mathrm{f}}\right] -
      \rho_0^\mathrm{f}\mathrm{\small{GRAD}}\mu^\mathrm{f}\right),
\label{fluidflux}
\end{equation}

\noindent where $\bD^\mathrm{f}$ is the positive semi-definite
mobility of the fluid, and isothermal conditions are assumed in order to
approximate the physiological ones. Experimentally determined
transport coefficients (e.g. for mouse tail skin \citep{Swartzetal:99}
and rabbit Achilles tendons \citep{Hanetal:2000}) are used for the
fluid mobility values. The terms in the parenthesis on the right
hand-side of \mbox{Equation (\ref{fluidflux})} sum to give the
total driving force for transport. The first term is the contribution
due to gravitational acceleration. In order to maintain physiological
relevance, this term has been neglected in the following
treatment. The second term arises from stress divergence; for an ideal
fluid, it reduces to a pressure gradient, thereby specifying that the
fluid moves down a compressive pressure gradient, which is Darcy's
Law. The third term is the gradient of the chemical
potential, $\mu^\mathrm{f} = e^\mathrm{f} - \theta \eta^\mathrm{f}$,
where $e^\mathrm{f}$ is the mass-specific internal energy, $\theta$ is
temperature and $\eta^\mathrm{f}$ is the mass-specific entropy. The
entropy gradient included in this term results in 
classical Fickean diffusion if only mixing entropy exists, as
discussed in the following section. For a
detailed derivation and discussion of \mbox{Equation
(\ref{fluidflux})}, the reader is directed to \citet{growthpaper}.




\subsubsection{Saturation and Fickean diffusion of the fluid}
\label{fick}

\begin{figure}
\centering
\psfrag{A}{\small A}
\psfrag{B}{\small B}
\psfrag{C}{\small C}
\psfrag{D}{\small Vacant space}
\psfrag{E}{\small Filled space}
\includegraphics[width=7.50cm]{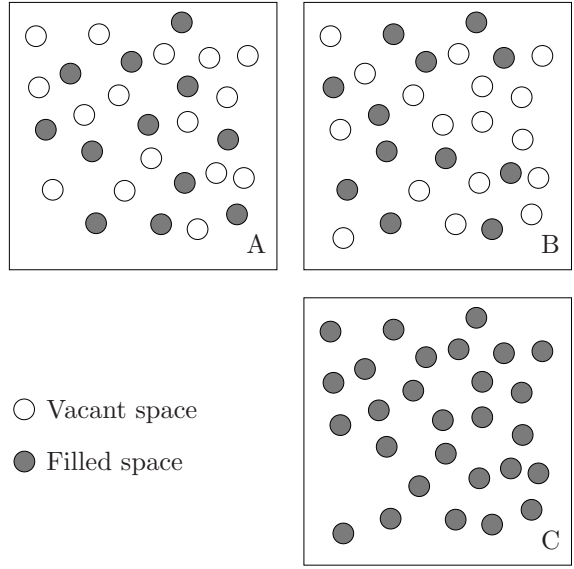}
\caption{Depicted at a microscopic scale, only unsaturated tissues A
  and B can undergo Fickean diffusion of the fluid. C is saturated.}
\label{fick_fig}
\end{figure}

As depicted in Figure~\ref{fick_fig}, only when pores are unsaturated
are there multiple configurations 
available to the fluid molecules at a fixed fluid concentration.  This
leads to a non-zero mixing entropy. In contrast, if saturated, there
is a single available configuration (degeneracy), resulting in
zero mixing entropy. Consequently, Fickean diffusion, which arises
from the gradient of mixing entropy can exist only in the unsaturated
case. However, even a saturated pore structure can demonstrate
concentration gradient-dependent mass transport phenomenologically: The
fluid stress depends on fluid concentration (see Equation (\ref{Pf})),
and fluid stress gradient-driven flux appears as a concentration
gradient-driven flux.

The saturation dependence of Fickean diffusion is modelled by using
the measure of saturation introduced in Section~\ref{satswel}. We
rewrite the chemical potential as

\begin{eqnarray}
\mu^\mathrm{f} &=&  
e^\mathrm{f} - \theta\eta^\mathrm{f},\nonumber\\
\eta^\mathrm{f} &\to& 0, \quad \mbox{as}\, \tilde{v}^\mathrm{f} +
\tilde{v}^\mathrm{c} \to 1.
\label{fickeanmobility}
\end{eqnarray}

\noindent It is again important to note that under physiological
conditions, soft tissues are fully saturated by fluid, and it is
appropriate to set $\mu^\mathrm{f} = e^\mathrm{f}$.

\subsubsection{Transport of solute species}
\label{solutespecies}

The dissolved solute species, denoted by s, undergo long range
transport primarily by being 
advected by the fluid. In addition to this, they undergo diffusive
transport relative to the fluid. This motivates an additional velocity
split of the form $\bV^s=\widetilde{\bV^\mathrm{s}}+\bV^\mathrm{f}$,
where $\widetilde{\bV^\mathrm{s}}$ denotes the velocity of the solute
relative to the fluid. The constitutive relation for the corresponding
flux, denoted by $\widetilde{\bM^s}$, has the following form, similar
to Equation (\ref {fluidflux}) defined for the fluid flux.


\begin{equation}
\widetilde{\bM^\mathrm{s}} = \bD^\mathrm{s}\left(
- \rho^\mathrm{s}_0\mathrm{\small{GRAD}}\left[e^\mathrm{s} -
 \theta\eta^\mathrm{s}\right]\right),
\label{soluteflux}
\end{equation}

\noindent where $\bD^\mathrm{s}$ is the positive semi-definite
mobility of the solute relative to the fluid, and again, isothermal
conditions are assumed to approximate the physiological
ones. Following Section \ref{bomom} there are no stress-dependent
contributions to $\widetilde{\bM^\mathrm{s}}$.

\subsubsection{Frame invariance and the contribution from acceleration}
\label{dropping_accn}

In our earlier treatment \citep{growthpaper}, the constitutive
relation for the fluid flux had a driving force contribution arising
from the acceleration of the solid phase,
$-\rho_0^\mathrm{f}\bF^{\mathrm{T}}\frac{\partial \bV}{\partial t}$.
This term, being motivated by the reduced dissipation inequality, does
not violate the Second Law and supports an intuitive understanding
that the acceleration of the solid skeleton in one direction must result in
an inertial driving force on the fluid in the opposite
direction. However, as defined, this acceleration is obtained by the
time differentiation of kinematic quantities,\footnote{And not in terms
of acceleration {\em relative to fixed stars} for e.g., as discussed
in \cite[][Page 43]{TruesdellNoll:65}.} and does not transform in a
frame-indifferent manner. Unlike the superficially similar term
arising from the gravity vector,\footnote{Where every observer has an
implicit knowledge of the directionality of the field relative to a
fixed frame, allowing it to transform objectively. Specifically, under
a time-dependent rigid body motion imposed on the current
configuration carrying $\bx$ to $\bx^+ = \bc(t) + \bQ(t)\bx$, where
$\bc(t) \in \mathbb{R}^3$ and $\bQ(t) \in \mbox{SO}(3)$, it is
understood that the acceleration due to gravity in the transformed
frame is $\bg^+ = \bQ^\mathrm{T}\bg$ and is therefore
frame-invariant. However, $\ba^+ = \ddot{\bc} + 2\dot{\bQ}\bv +
\ddot{\bQ}\bx + \bQ\ba$ , and is therefore not frame-invariant.} the
acceleration 
term presents an improper dependence on the frame of the
observer. Thus, its use in constitutive relations is inappropriate,
and the term has been dropped in \mbox{Equation (\ref{fluidflux})}.

\subsubsection{Incompressible fluid in a porous solid}
\label{incompfluid}

Upon incorporation of the additional velocity split,
$\bV^\mathrm{s}=\widetilde{\bV^\mathrm{s}}+\bV^f$, described in
Section~\ref{solutespecies}, the resulting mass transport equation
(\ref{massbalcurr}) for the solute species is

\begin{equation}
\frac{\mathrm{d}\rho^\mathrm{s}}{\mathrm{d}t} = \pi^\mathrm{s}-
\mathrm{div} \left[\widetilde{\bm^\mathrm{s}}+
\frac{\rho^\mathrm{s}}{\rho^f}\bm^f\right] - \rho^\mathrm{s}
\mathrm{div}[\bv].
\label{massbalcurrsol}
\end{equation}

\noindent In the hyperbolic limit, where advection dominates, spatial
oscillations emerge in numerical solutions of this equation
\citep{Brooks:82,Paper6}. However, the form in which the equation is
obtained is not in standard advection-diffusion form, and therefore is
not amenable to the application of standard stabilisation techniques
\citep{Paper6}. In part, this is because although the (near)
incompressibility of the fluid phase is embedded in the balance of
linear momentum via the fluid stress, it has not yet been explicitly
incorporated into the transport equations. This section proceeds to
impose the fluid incompressibility condition and deduces implications
for the solute mass transport equation, including a crucial
simplification allowing for its straightforward numerical
stabilisation.

From \mbox{Equation (\ref{massbalcurr})}, the local form of the
balance of mass for the fluid species (recalling that
$\Pi^\mathrm{f}=0$) in the current configuration is

\begin{equation}
\frac{\mathrm{d}\rho^f}{\mathrm{d}t} = - \mathrm{div}\left[\bm^f\right]
- \rho^f \mathrm{div}\left[\bv\right].
\label{fluidtransporteqn}
\end{equation}

\noindent In order to impose the incompressibility of the fluid, we
first denote by $\rho_{0_{\mathrm{ini}}}^{f}$ the {\sl initial} value
of the fluid reference concentration. Recall that the fluid
concentration with respect to the reference configuration evolves in
time; $\rho^\mathrm{f}_0 = \rho^\mathrm{f}_0(\bX,t)$. Therefore we can
precisely, and non-trivially, define
$\rho_{0_{\mathrm{ini}}}^{\mathrm{f}}(\bX)$   

\begin{equation}
\begin{split}
\rho_{0}^{\mathrm{f}}(\bX,0)
                   &=:\rho_{0_{\mathrm{ini}}}^{\mathrm{f}}(\bX)\\ 
                   &=\rho_{\mathrm{ini}}^{\mathrm{f}}(\bx\circ\Bvarphi)
                   J(\bX, t)\\ &=\frac{\rho^{\mathrm{f}}
                   (\bx\circ\Bvarphi,t)} {J^{f_\mathrm{g}}(\bX,t)}
                   J(\bX,t)\\ &=\rho^{\mathrm{f}} (\bx\circ\Bvarphi,t)
                   \cancelto{\approx 1\ \forall\ t}
                   {J^{f_\mathrm{e}}}(\bX,t).\\
\label{incompderiv}
\end{split}
\end{equation}

In (\ref{incompderiv}), $J := \mathrm{det}(\bF)$ and $J^{f_\mathrm{g}} :=
\mathrm{det}(\bF^{\mathrm{g}^{\mathrm{f}}})$. The quantity
$\rho_{\mathrm{ini}}^{\mathrm{f}}$ is defined by the right hand-sides
of the first and second lines of (\ref{incompderiv}). To follow the
argument, consider, momentarily, a \emph{compressible} fluid. If the
current concentration, $\rho^\mathrm{f}$, changes due to elastic
deformation of the fluid and by transport, then
$\rho_{\mathrm{ini}}^{\mathrm{f}}$ as defined is not a
physically-realized fluid concentration. It bears a purely
mathematical relation to the current concentration, $\rho^\mathrm{f}$,
since the latter quantity represents the effect of deformation of a
tissue point as well as change in mass due to transport at that
point. If the contribution due to mass change at a point is scaled
out of $\rho^\mathrm{f}$ the quotient is identical to the result of
dividing $\rho_{0_{\mathrm{ini}}}^{\mathrm{f}}$ by the deformation
only. This is expressed in the relation between the right hand-sides
of the second and third lines of (\ref{incompderiv}). The elastic
component of fluid volume change in a pore is $J^{f_\mathrm{e}} :=
\mathrm{det}(\bF^{\mathrm{e}^{\mathrm{f}}})$, which appears  in the third
line of (\ref{incompderiv}) via the preceding arguments. Clearly then,
for a fluid demonstrating near incompressibility intrinsically (i.e.,
the true density is nearly constant), we have $J^{f_\mathrm{e}}
\approx 1$ as indicated. Equation (\ref{incompderiv}) therefore shows
that for a nearly incompressible fluid occupying the pores of a
tissue, if we further assume that the pore structure deforms as the
solid collagenous skeleton, $\rho_0^\mathrm{f}(\bX,0) \approx
\rho^\mathrm{f}(\bx\circ\Bvarphi,t)$. The fluid concentration as
measured in the current configuration is approximately constant in
space and time. This allows us to write,

\begin{equation}
\frac{\partial} {\partial t}\left(
\rho_{0_{\mathrm{ini}}}^{f}(\bX) \right) \equiv 0 \Rightarrow
\frac{\partial} {\partial t}\left(\rho^{f} (\bx\circ\Bvarphi,t)
\right)\Big\vert_{\bX} = 0,
\end{equation}

\noindent which is the hidden implication of our assumption of a
homogeneous deformation, i.e., $\bF$ is the deformation gradient of
solid collagen and the pore spaces.  This leads to $\frac{\mathrm{d}
  \rho^f} {\mathrm{d} 
  t}=0$.\footnote{Which results in a very large pressure gradient
  driven flux due to incompressibility. } We
therefore proceed to treat our fluid mass transport at steady
state. Rewriting the flux $\bm^{\mathrm{f}}$ from \mbox{Equation
(\ref{fluidtransporteqn})} as the product $\rho^{\mathrm{f}}
\bv^{\mathrm{f}}$ and using the result derived above,
\begin{equation}
\begin{split}
0 &= \left. \frac{\partial \rho^f}{\partial t} \right|_{\bX}\\ &=
-\mathrm{div}\left[\rho^f \bv^{f}\right] - \rho^f
\mathrm{div}\left[\bv\right].
\end{split}
\label{incomprimpl}
\end{equation}

\noindent Returning to (\ref{massbalcurrsol}) with this result,

\begin{equation}
\begin{split}
\frac{\mathrm{d}\rho^\mathrm{s}}{\mathrm{d}t} &= \pi^\mathrm{s}-
\mathrm{div} \left[\widetilde{\bm^\mathrm{s}}+
\frac{\rho^\mathrm{s}}{\rho^f}\bm^f\right] - \rho^\mathrm{s}
\mathrm{div}[\bv]\\ &= \frac{\rho^\mathrm{s}}{\rho^f}\left(\cancelto{0}
{-\mathrm{div}\left[\rho^f \bv^{f}\right] - \rho^f
\mathrm{div}[\bv]}\right)\\ &\quad{} + \pi^\mathrm{s} -
\mathrm{div}\left[\widetilde{\bm^\mathrm{s}}\right]
-\bm^f\cdot\mathrm{grad}\left[\frac{ \rho^\mathrm{s}}{\rho^f}\right].\\
\end{split}
\end{equation}

Thus, using the incompressibility condition (\ref{incomprimpl}), we
get the simplified form of the balance of mass for an arbitrary solute
species, s,

\begin{equation}
\frac{\mathrm{d}\rho^\mathrm{s}}{\mathrm{d}t}=\pi^\mathrm{s} -
\mathrm{div}\left[\widetilde{\bm^\mathrm{s}}\right] -
\frac{\bm^f\cdot\mathrm{grad}\left[\rho^\mathrm{s}\right]}{\rho^f} +
\frac{\rho^\mathrm{s} \bm^f \cdot \mathrm{grad}\left[\rho^f\right]}
{\rho^{f^2}}.
\label{stdform}
\end{equation}

\noindent Using the pushed-forward form of (\ref{soluteflux}), this is
now in standard advection-diffusion form, 

\begin{equation}
\begin{split}
& \frac{\mathrm{d}\rho^\mathrm{s}}{\mathrm{d}t} - \underbrace{
 \mathrm{div}\left[\bar{\bD^\mathrm{s}} \mathrm{grad}
 \left[ \rho^\mathrm{s}\right]\right]}_\text{Diffusion term}
 - \underbrace{\pi^\mathrm{s}}_\text{Source term} =
\\ \ & - \underbrace{\frac{\bm^f\cdot\mathrm{grad}\left[\rho^\mathrm{s}\right]}{\rho^f}}
_\text{Advection term} +
\underbrace{ \frac{\rho^\mathrm{s} \bm^f \cdot \mathrm{grad}\left[\rho^f\right]}
     {\rho^{f^2}},}_\text{Additional, $\rho^\mathrm{s}$-dependent source term}
\end{split}
\label{morestdform}
\end{equation}

\noindent where $\bar{\bD^\mathrm{s}}$ is a positive semi-definite
diffusivity, $\bm^{f}/\rho^{f}$ is the advective velocity, and $
\pi^\mathrm{s}$ is the volumetric source term. This form is well
suited for stabilisation schemes such as the streamline upwind
Petrov-Galerkin (SUPG) method (see, for e.g., \cite{Paper6}), described
briefly below, which limit spatial oscillations otherwise observed
when the element {\em Peclet number} is large.

\subsubsection{Stabilisation of the simplified solute transport
equation}\label{solutetranspstab}

In weak form, the SUPG-stabilised method for
Equation~(\ref{morestdform}) is

\begin{equation}
\begin{split}
&\int_{\Omega} w^{\mathrm{h}} \left(
  \frac{\mathrm{d}\rho^{\mathrm{s}^{h}}}{\mathrm{d}t} +
  \bm^f\cdot\mathrm{grad}\left[\frac{
      \rho^{\mathrm{s}^{h}}}{\rho^f}\right] \right)
  d\Omega\\ &+\int_{\Omega} \left( \mathrm{grad}
  \left[w^{\mathrm{h}}\right] \cdot \bar{\bD^\mathrm{s}} \mathrm{grad}
  \left[ \rho^{\mathrm{s}^{h}}\right] \right)\ d\Omega\\ +&
  \sum_{\mathrm{e}=1}^{\mathrm{n_{el}}} \int_{\Omega_{\mathrm{e}}}
  \tau \frac{\bm^{f}}{\rho^f} \cdot \mathrm{grad} \left[w^{\mathrm{h}}\right] \left(
  \frac{\mathrm{d}\rho^{\mathrm{s}^{h}}}{\mathrm{d}t} +
  \bm^f\cdot\mathrm{grad}\left[\frac{
      \rho^{\mathrm{s}^{h}}}{\rho^f}\right] \right) \ d\Omega\\ -&
  \sum_{\mathrm{e}=1}^{\mathrm{n_{el}}} \int_{\Omega_{\mathrm{e}}}
  \tau \frac{\bm^{f}}{\rho^f} \cdot \mathrm{grad} \left[w^{\mathrm{h}}\right]
  \left(\mathrm{div}\left[\bar{\bD^\mathrm{s}}\ \mathrm{grad} \left[
      \rho^{\mathrm{s}^{h}}\right]\right]\right) \ d\Omega\\ = &
  \int_{\Omega} w^{\mathrm{h}} \pi^\mathrm{s} \ d\Omega +
  \int_{\Gamma_{\mathrm{h}}} w^{\mathrm{h}} h \ d\Gamma\\ +&
  \sum_{\mathrm{e}=1}^{\mathrm{n_{el}}} \int_{\Omega_{\mathrm{e}}}
  \tau \frac{\bm^{f}}{\rho^f} \cdot \mathrm{grad} \left[w^{\mathrm{h}}\right]
  \pi^\mathrm{s} \ d\Omega,
\label{stabilizedmassbal}
\end{split}
\end{equation}

\noindent where quantities with the superscript $\mathrm{h}$ represent
finite-di\-men\-sion\-al approximations of infinite-dimensional field
variables, $\Gamma_{\mathrm{h}}$ is the Neumann boundary, and 
this equation introduces a numerical stabilisation parameter $\tau$,
which we have calculated from the $\mathrm{L}_{2}$ norms of element
level matrices, as described in
\cite{tezduyarsupg}.

\subsection{Nature of the sources}
\label{sources}

There exists a large body of literature,
\citep{CowinHegedus:76,EpsteinMaugin:2000,AmbrosiMollica:2002}, that
addresses growth in biological tissue mainly based upon a single
species undergoing transport and production/annihilation. However,
when chemistry is accounted for, it is apparent that growth depends on
cascades of complex biochemical reactions involving several species,
and additionally involves intimate coupling between mass transfer,
biochemistry and mechanics. An example of this chemo-mechanical
coupling is described in \cite{Provenzanoetal:2003}.

The modelling approach followed in this work is to select appropriate
functional forms of the source terms for collagen, $\Pi^{\mathrm{c}}$,
and the solutes, $\Pi^{\mathrm{s}}$, that abstract the complexity of
the biochemistry. In our earlier exposition \citep{growthpaper}, we
used simple first order chemical kinetics to define
$\Pi^{\mathrm{c}}$. Other forms, which have been studied in the
literature, can be used: 

(\romannumeral 1) {\em Michaelis-Menten}
enzyme kinetics (see, for 
e.g., \cite{Sengersetal:2004}), which involves a two-step reaction
with the collagen and solute production terms given by

\begin{equation}
\Pi^\mathrm{s} =
    \frac{-(k_{\mathrm{max}}\rho^{\mathrm{s}})}
    {(\rho^{\mathrm{s}}_m+\rho^{\mathrm{s}})}
    \rho_{\mathrm{cell}}, \quad\Pi^\mathrm{c} = -\Pi^\mathrm{s},
\label{enzymekineticseq}
\end{equation}

\noindent where $\rho_{\mathrm{cell}}$ is the concentration of
fibroblasts, $k_{\mathrm{max}}$ is the maximum value of the solute
production reaction rate constant, and $\rho^{\mathrm{s}}_m$ is half
the solute concentration corresponding to $k_{\mathrm{max}}$. For
details on the chemistry modelled by the Michaelis-Menten model, see,
for e.g., \citet{sbromadill}.

(\romannumeral 2) {\em Strain energy-dependent} sources that
induce growth at a point when the 
energy density deviates from a reference value. An example of source terms
of this form was
originally proposed in the context of bone growth 
\citep{HarriganHamilton:93}. We are not aware of studies that have
developed similar functional forms for soft tissue, and therefore have
adapted this example from the bone growth literature, recognizing that
this topic is in need of further study. Suitably weighted by a
relative concentration ratio, and written for collagen, this source
term has the form 

\begin{equation}
\Pi^\mathrm{c} =
\left(\frac{\rho^\mathrm{c}_0}{\rho^\mathrm{c}_{0_\mathrm{ini}}}\right)^{-m}
\psi_{\mathrm{F}}-\psi_{\mathrm{F}}^*,
\label{strainsrc}
\end{equation}

\noindent where $\psi_{\mathrm{F}}$ is the mass-specific strain energy
function, and $\psi_{\mathrm{F}}^*$ is a reference value of this
strain energy density. Equation (\ref{strainsrc}) models collagen
production when the strain energy density (weighted by a concentration
ratio) at a point exceeds this reference value, and models 
annihilation otherwise.

\section{Numerical examples}
\label{numericalimplementation}

\begin{figure}
\centering
  \includegraphics[width=7.50cm]{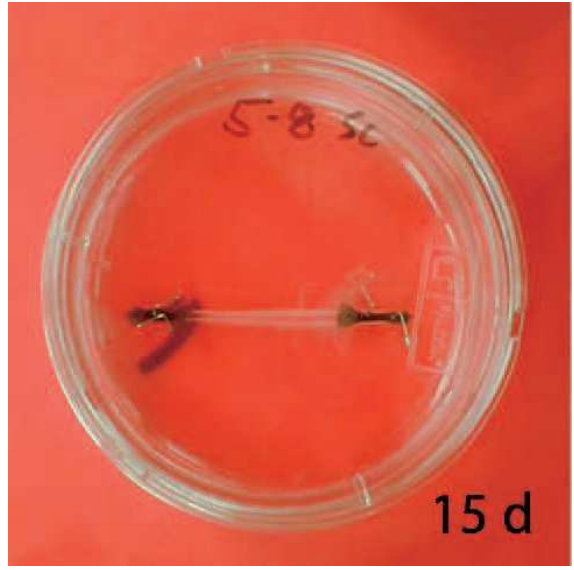}
\caption{Engineered tendon constructs. See \citet{Calve:04} for
  details.} 
\label{engconst}
\end{figure}

\begin{figure}[ht]
  \centering
  \psfrag{D}{\small$1.1284$ mm}
  \psfrag{H}{\small$12.0$ mm}
  {\includegraphics[width=5cm]{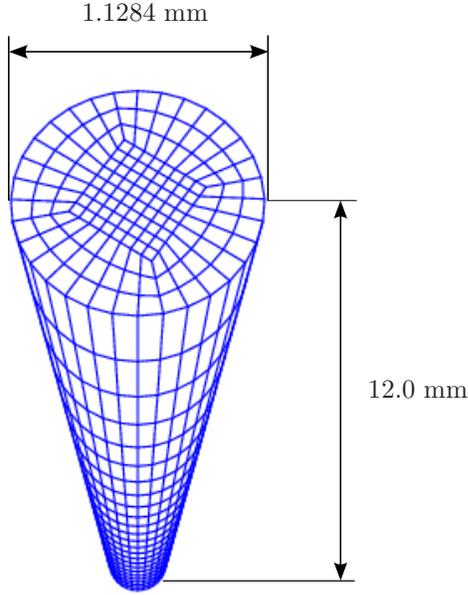}}
  \caption{The finite element mesh used in the computations.}
  \label{egmesh}
\end{figure}

The theory presented in the preceding sections results in a system of
non-linear, coupled partial differential equations. A finite element
formulation employing a staggered scheme based upon operator splits
\cite{Armero-poroplasticity:99,Garikipatiox2:01} has been implemented
in {\tt FEAP} \citep{feapmanual} to solve the coupled problem. As an
example, in the biphasic problem involving solid and fluid
phases only, the basic solution scheme involves keeping the displacement
field fixed while solving for the concentration fields using the mass
transport equations. The resulting concentration fields are then fixed
to solve the mechanics problem. This procedure is repeated until the
resulting fields satisfy the differential equations within a specified
numerical tolerance.

The following examples aim to demonstrate the mathematical formulation
and aspects of the coupled phenomena as the tissue grows. The model
geometry, based on our engineered tendon constructs (see
Figure~\ref{engconst} and \citet{Calve:04}), is a cylinder 12 mm in length and 1 mm$^2$ in
cross-sectional area. The corresponding finite element mesh using
hexahedral elements, is shown in Figure~\ref{egmesh}.

The following numerical examples involve solution of a common set of
partial differential equations. The constutive models, however, vary as we
demonstrate the behaviour engendered by the many modelling
assumptions discussed in the paper. The balance of linear momentum
that we solve is (\ref{linearmombalance}) summed for $\iota =
\mathrm{c,f}$, with the constraint in (\ref{qrelation}) imposed. The
absence of significant acceleration in the problems under 
consideration allows us to solve the balance of linear momentum   
quasi-statically. The fluid mass balance equation is solved in the current
configuration, i.e. (\ref{massbalcurr}) for $\iota = \mathrm{f}$, but 
mass balance for the solid collagenous phase is solved in the
reference configuration, i.e. (\ref{massbalance1}) for $\iota =
\mathrm{c}$.  Mass balance for the solute is also solved in the
current configuration, but using the stabilized scheme in weak
form (\ref{stabilizedmassbal}). The Backward 
Euler algorithm is used for all mass transport equations. The
constitutive relation for the solid collagen follows
(\ref{wlcmeq}). The constitutive relation for the fluid stress follows
(\ref{Pf}) with 
\begin{equation}
h(\rho^\mathrm{f}) =
\frac{1}{2}\kappa^\mathrm{f}\left(\frac{\rho_{0_\mathrm{ini}}^\mathrm{f}}{\rho^\mathrm{f}}
- 1\right)^2,
\end{equation}

\noindent where $\kappa^\mathrm{f}$ is the fluid bulk modulus. The
tissue is modelled as being fluid saturated in $\Omega_t$ at $t = 0$,
i.e. (\ref{saturation}$_1$) holds with
$\rho^\mathrm{f}_{0_\mathrm{sat}} =
\rho^\mathrm{f}_{0_\mathrm{ini}}$. However, the tissue is allowed to 
become unsaturated in $\Omega_t$ for $t > 0$ due to void formation. Then,
the conditions set out in (\ref{cavitation})
apply. The chemical potential is then given by
(\ref{fickeanmobility}). The numerical examples that follow discuss further
specialization of the constitutive relations to other cases discussed
in the preceding sections. The numerical values of parameters\footnote{The
  mobility tensor reported in Table~\ref{parameters} is an
  order-of-magnitude estimate 
  recalculated from \citet{Hanetal:2000} to correspond to the 
  mobility used in this paper. These authors reported a mean value of
  $0.927\times 10^{-14}$ s, with a range of $1.14\times
  10^{-14}--0.58\times 10^{-14}$ s in terms of the mobility used
  here. Theirs is the mobility parallel to the fiber direction in
  Rabbit Achilles tendon. Our usage of it is as an isotropic
  mobility. Using anisotropic mobilities, or different values from the
  reported range changes the result
  quantitatively, but not qualitatively.}  that
have been used appear in Table \ref{parameters}.

Non-linear projection methods \citep{simotaylorpister:85} are used to treat the
near-incompressibility imposed by the fluid. Mixed methods, as described
in \cite{Garikipatiox2:01}, are used for stress (and strain) gradient
driven fluxes.

The initial and boundary
conditions have been chosen in order to model a few common mechanical and
chemical interventions on engineered tissue. However, we will not
attempt detailed descriptions of experiments, choosing to focus
instead on results that can be directly related to the models. A more
detailed comparison with experiments is forthcoming in a separate
communication. 

\subsection{A multiphasic problem based on enzyme-kinetics}
\label{enzyme_kinetics_eg}

\begin{table}
\centering
\begin{tabular}{lll}
\hline
\multicolumn{1}{c}{Parameter (Symbol)} & Value & Units\\
\hline
Chain density ($N$) & $7\times 10^{21}$ & $\mathrm{m}^{-3}$\\
Temperature ($\theta$)  & $310.6$ & K\\
Persistence length ($A$) & $2.10$ & --\\
Fully-stretched length ($L$) & $2.125$ & --\\
Unit cell axes ($a,\;b,\;c$) & $1.95,\;1.95,\;2.43$ & --\\
Bulk compressibility factors ($\gamma,\;\beta$) & $1000,\; 4.5$ & --\\
Fluid bulk modulus ($\kappa^f$) & $1$ & GPa\\
Fluid mobility tensor ($D^\mathrm{f}_{ij} = D^\mathrm{f}\delta_{ij}$) & $1\times 10^{-14}$
&s\\
Fibroblast concentration ($\rho_{\mathrm{cell}}$) & 0.2 &
kg.m$^{-3}$\\
Max. reaction rate ($k_{\mathrm{max}} = 5$) & 5 & s$^{-1}$\\
Max. solute concentration ($\rho^{\mathrm{s}}_m$) & 0.2 &
kg.m$^{-3}$\\
Solute diffusivity ($\bD^\mathrm{s}$) & $1\times 10^{-9}$ &  m$^{-2}$s\\
\hline
\end{tabular}
\caption{Material parameters used in the analysis.}
\label{parameters}
\end{table}

\begin{figure}
\centering
\includegraphics[angle=270,width=7.50cm]{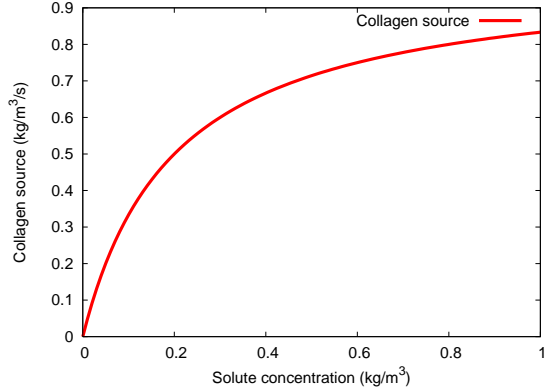}
\caption{Variation of the collagen source term (kg.m$^{-3}$.s$^{-1}$)
  with solute concentration (kg.m$^{-3}$).}
\label{eg3menten}
\end{figure}

This first example can be viewed as a model for localised, bolus
delivery of regulatory chemicals to the tendon while accounting for
mechanical (stress) effects. A single solute species\footnote{Here, we
  envision the solute to be a protein playing 
  an essential role in growth by catalysing underlying biochemical
  reactions. An important example of this is a family of proteins,
  TGF$\beta$, which is a multi-functional peptide that controls numerous
  functions of many cell types \citep{Alberts:02}.} is considered,
denoted by s, and 
a uniform distribution of fibroblasts that are characterised by their
cell concentration, $\rho_{\mathrm{cell}}$. Both, Fickean diffusion of
the solute, and stress gradient driven fluid flow are incorporated in this
illustration. We use 
Michaelis-Menten enzyme kinetics [Equation~(\ref{enzymekineticseq})]
to determine the rates of solute consumption and collagen production
as a function of solute concentration. This non-linear relationship for
our choice of parameters is visualised in
Figure~\ref{eg3menten}. Here, the fluid phase does not  take part in
reactions, and hence $\Pi^\mathrm{f}=0$. 

\begin{figure}[ht]
  \centering
  \psfrag{M}{\small $\bN\cdot\bM^f$}
  \psfrag{P}{\small ${\tiny\quad }~\bu$}
         {\includegraphics[width=5.0cm]{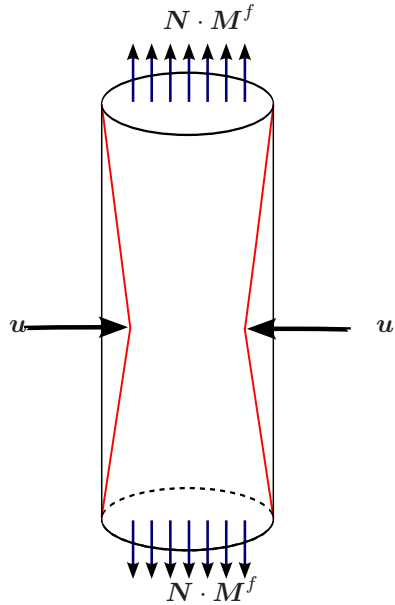}}
	 \caption{Constrictive load applied to tendon immersed in a
	 bath.} 
	 \label{constrictload}
\end{figure}

The tendon immersed in the bath is subjected to a constrictive
radial load, such
as would be imposed upon  manipulating it with a set of tweezers, as depicted in
Figure~\ref{constrictload}. The
maximum strain in the radial 
direction---experienced half-way through the height of the tendon---is
10\%. The applied strain in the radial direction decreases linearly
with distance from the central plane, and vanishes at the top and
bottom surfaces of the tendon.

\begin{figure}
\centering
\includegraphics[width=7.50cm]{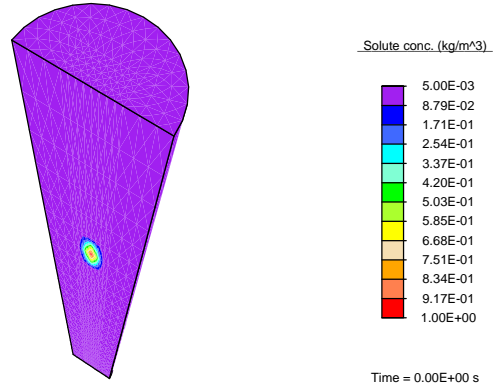}
\caption{The solute concentration (kg.m$^{-3}$) initially.}
\label{eg3ini}
\end{figure}

The 
initial collagen concentration and the initial fluid concentration are
both 500~kg.m$^{-3}$ at every point in the tendon, and the fluid
concentration in the bath is 500~kg.m$^{-3}$. In
addition, a solute-rich bulb of radius 0.15~mm is introduced with
its centre on the axis of the tendon and situated 3~mm below the upper
circular face of the tendon. The initial solute concentration is
0.05~kg.m$^{-3}$ at all other points in the tendon, and increases
linearly with decreasing radius in this bulb to 1~kg.m$^{-3}$ at its
centre (see Figure~\ref{eg3ini}). The
parameters used are listed in Table~\ref{parameters}, and are relevant
to tendons.

\begin{figure}
\centering
\includegraphics[width=7.50cm]{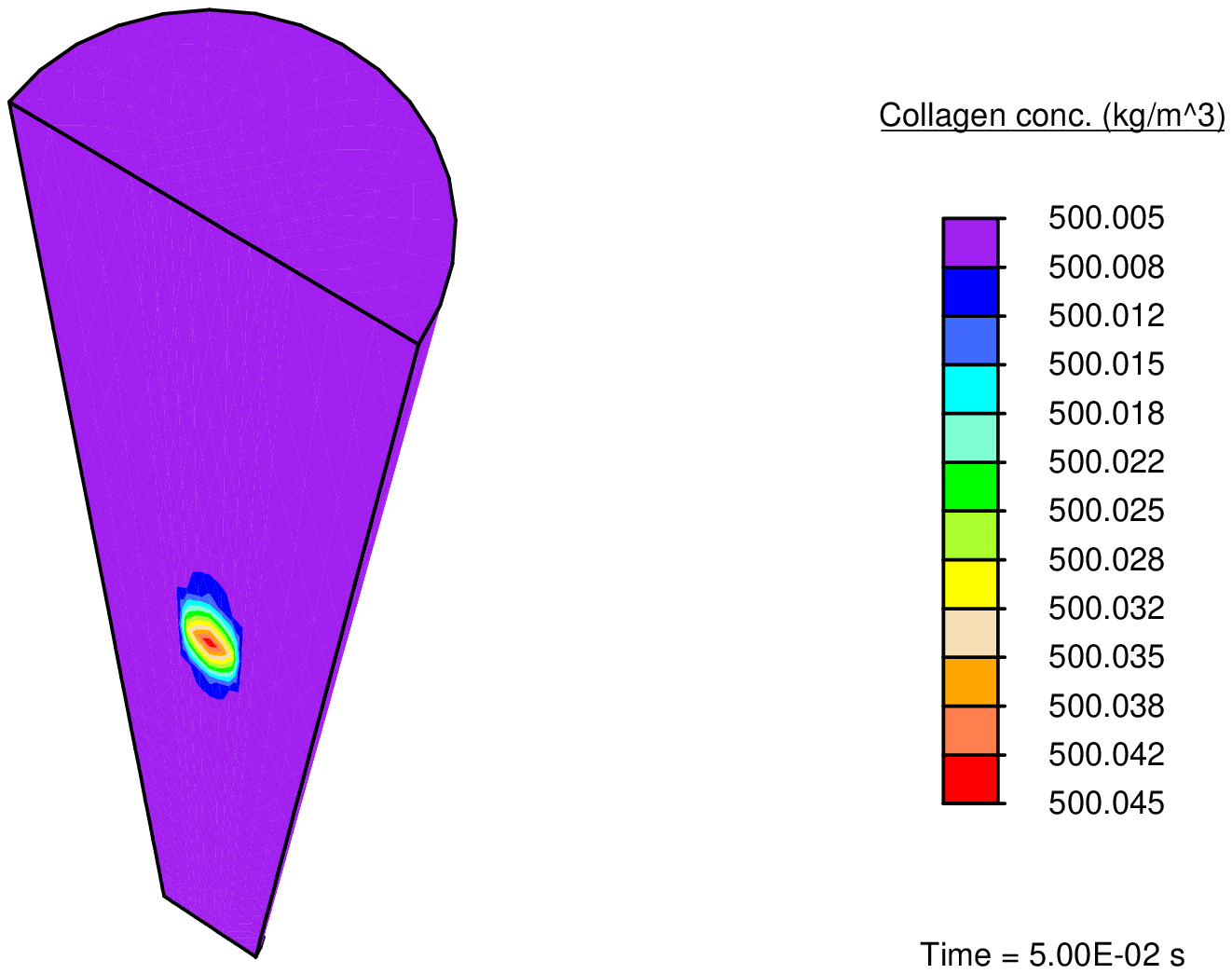}
\caption{The collagen concentration (kg.m$^{-3}$) at time
  $t=5\times10^{-2}$~s.}
\label{eg3conc}
\end{figure}

The aim of this example is to compare the influences upon solute
transport from two mechanisms: Fluid stress gradient-driven transport,
arising from the applied constrictive load, and solute concentration
gradient-driven transport. These mechanisms have both been implicated
in nutrient supply to cells in soft tissue. The results of this
numerical example demonstrate that because the magnitude of the fluid mobility
for stress gradient driven transport is orders of magnitude
smaller than the diffusion coefficient for the solute through the
fluid, there is relatively only a small stress gradient driven flux,
and the transport of the solute is diffusion dominated. As a result,
the solute diffuses locally, but displays no
observable advection along the fluid. As the diffusion-driven solute
concentration in a region increases, the enzyme-kinetics model
results in a small source term for collagen production, and we observe nominal
growth. Figure~\ref{eg3conc} shows the collagen concentration at an
early time, $t=5\times10^{-2}$~s.

This example incorporates all of the theory discussed in the
paper. However, it is a valuable exercise in
modelling to simplify the boundary value problem, and supress some of
the coupled phenomena in order to gain a better understanding of some
effcts. This is the approach followed in the next two numerical
examples. The detailed transport and mechanics induced by the
constrictive radial load are discussed first in Section~\ref{pinching}. 

\subsection{Examples exploring the biphasic nature of porous soft tissue}
\label{firstorder}

In these calculations, only two phases---fluid and collagen---are
included for the mass transport and mechanics. The parameters used in the analysis
are presented in Table~\ref{parameters}. 


\subsubsection{The tendon under constriction}
\label{pinching}

In this example, the tendon immersed in a bath is subjected to the same
constrictive radial load as in Section~\ref{enzyme_kinetics_eg}. Since
that example demonstrated an insignificant amount of local collagen production
over this time scale, we have simplified the
problem by setting the source term $\Pi^\mathrm{c} = 0$. The total
duration of the simulation is 
10~s, and the radial strain is applied as a displacement boundary
condition, increasing linearly from no strain initially to the maximum
strain at time $t = 1~\mathrm{s}$. Again,
both the initial collagen 
concentration and the initial fluid concentration are 500~kg.m$^{-3}$
at every point in the tendon. This tendon is exposed to a bath where
the fluid concentration is 500~kg.m$^{-3}$.

While solving the balance of momentum for the biphasic problem
of the solid collagen and a fluid phase, we currently treat the
tissue as a single entity and employ a summation of
Equation~(\ref{linearmombalance}) over both species. Additionally,
condition~(\ref{qrelation}) allows us to avoid constitutive
prescription of the momentum transfer terms between solid collagen and
fluid phases,
$\bq^\mathrm{c}$ and $\bq^\mathrm{f}$. This facilitates considerable
simplification of the 
problem, but such a treatment requires additional assumptions on the
detailed deformation of the constitutive phases of the tissue. An
explicit assumption we have drawn on thus far is the equality of
the deformation gradient of the solid collagen and pore spaces,
allowing us to use the deformation gradient
$\bF$, suitably decomposed to account for change in fluid
concentration, to model the fluid stress. This assumption and its 
consequences have been discussed in Sections \ref{bomass},
\ref{growthkinem}, \ref{satswel}, \ref{compfluid}, \ref{tensionfluid}
and \ref{incompfluid}. Since the imposition of a common deformation gradient
results in an upper bound for the 
effective stiffness of the tissue and magnitudes of the fluxes
established, we refer to it as the {\em upper bound model}. This
assumption plays a fundamental role in determining the fluid flux driven
by the fluid stress gradient.

\begin{figure}[ht]
  \centering
      {\includegraphics[width=7.5cm]{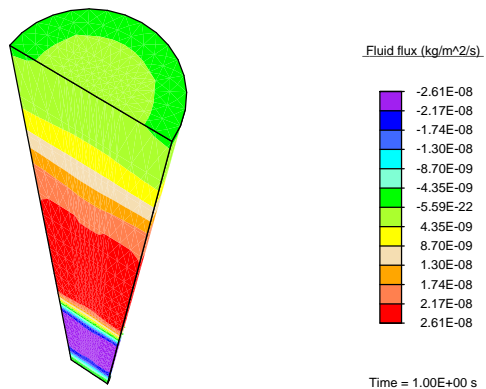}}
      \caption{{\em Upper bound} fluid flux (kg.m$^{-2}$.s$^{-1}$) in
        the vertical direction at time $t=1$~s.}
      \label{eg2flux}
\end{figure}

\begin{figure}[ht]
  \centering
      {\includegraphics[width=7.5cm]{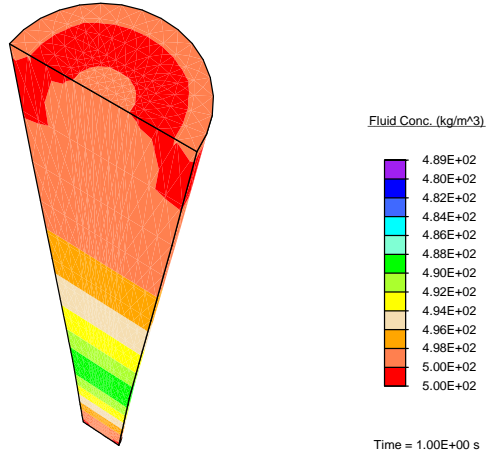}}
      \caption{Reference fluid concentration (kg.m$^{-3}$) at time
      $t=1$~s.}
      \label{eg2conc}
\end{figure}

\noindent For this upper bound model, Figure~\ref{eg2flux} shows the fluid flux in
the vertical direction at the final stage of the constriction phase of
the simulation, i.e. at time $t=1$~s. The flux values are positive
above the central plane, forcing fluid upward, and negative below,
forcing fluid fluid downward. This stress-gradient induced fluid flux
results in a reference concentration distribution of the fluid that is
higher near the top and bottom faces, as seen in Figure~\ref{eg2conc}.

As a result, these regions would have seen a higher production of
collagen, or preferential growth, in the presence of non-zero source
terms. As discussed in Section~\ref{curr-ref-mb}, the mass transport
equations are solved in the current configuration, where physical
boundary conditions can be set directly. The values reported in
Figure~\ref{eg2conc} are pulled back from the current
configuration. The current concentrations do not change for this
boundary value problem. 

Solving a problem of this nature in the
reference configuration using $\rho_0^\mathrm{f} = $ const. as the
boundary condition to represent immersion of the tendon in a fluid bath
yields non-physical results, such as an unbounded flow. This occurs
since the imposed strain gradient causes a stress gradient in the
fluid that does not decay. The imposed boundary condition on $\rho_0^\mathrm{f}$
prevents a redistribution of concentration that would have provided an
opposing, internal gradient of stress, which in turn would drive the
flux to vanish.

\begin{figure}[ht]
  \centering
      {\includegraphics[width=7.50cm]{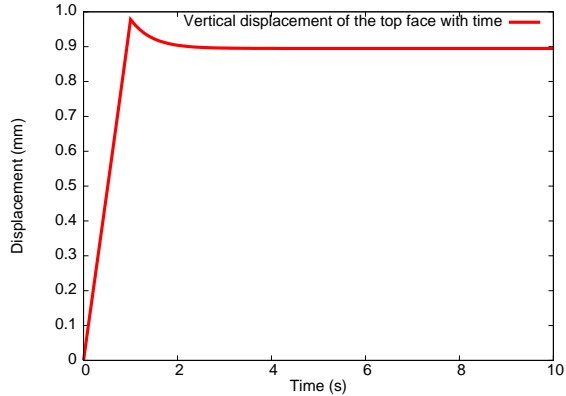}}
      \caption{Relaxation of the top face of the tendon after the
      constriction phase.}
      \label{topdisp}
\end{figure}

The tendon is held fixed in the radial direction after the
constriction phase. The applied stress sets up a pressure wave in the
fluid travelling toward the top and bottom faces. As the fluid leaves
these surfaces, we observe that the tendon relaxes. This is seen in
Figure~\ref{topdisp}, which plots the vertical displacement of the top
face with time, showing a decrease in height of the tendon after the
constriction phase. We keep the centre of the bottom face of the
tendon fixed.

\begin{figure}[ht]
  \centering
      {\includegraphics[width=7.50cm]{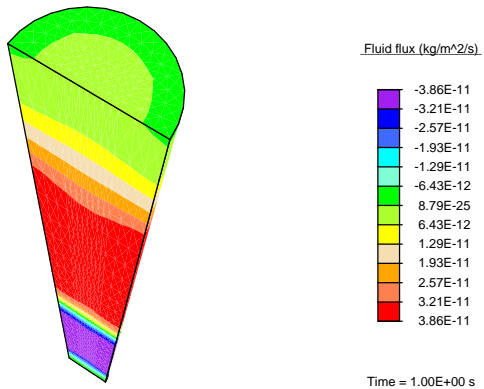}}
      \caption{{\em Lower bound} fluid flux (kg.m$^{-2}$.s$^{-1}$) in
        the vertical direction at time $t=1$~s.}
      \label{lowerbound}
\end{figure}

In order to define a range of the magnitude of fluid flux, we now
introduce the {\em lower bound model} (on effective stiffness of the
tissue and, consequently, the magnitude of the fluid flux). For this lower bound, we
replace the earlier strain homogenisation requirement with a stress
homogenisation requirement, {\em viz.} equating the hydrostatic stress
of the solid phase and the fluid pressure in the current
configuration:

\begin{equation}
p^{\mathrm{f}}=\frac{1}{3} \mathrm{\small{tr}}[\Bsigma^{c}],
\label{equalpr}
\end{equation}

\noindent where $p^{\mathrm{f}}$ is the fluid pressure in the current
configuration, $\mbox{\small{tr}[\textbullet]}$ is the trace operator, and
$\Bsigma^{c}=\frac{1}{\mathrm{J^{c}}} \bP^{\mathrm{c}}
\bF^{\mathrm{c}^{\mathrm{T}}}$ is the Cauchy stress of the solid. The
Cauchy stress of an ideal fluid can be defined from its current
pressure as \mbox{$\Bsigma^{f}= p^{\mathrm{f}} \bone$.}
Figure~{\ref{lowerbound}} reports the value of the vertical flux under
the lower bound modelling assumption, using boundary conditions identical to the
previous calculation at time $t=1$~s, the final stage of the
constriction phase of the simulation.

The fluid flux values reported in Figures~\ref{eg2flux} and
\ref{lowerbound} (corresponding to the upper and lower bound modelling
assumptions, respectively) are qualitatively similar, but differ by
about three orders of magnitude. This wide range points to the
importance of imposing the appropriate mechanical coupling model
between interacting phases. Note, however, that we have computed
bounds for the 
range of possible fluid flux values under the specified mechanical
loading. Recall, furthermore, that the example in Section
\ref{enzyme_kinetics_eg} used the upper bound model, and yet resulted
in no discernible advective solute transport. This suggests strongly
that, given the parameters in Table \ref{parameters}, convective
transport of nutrients in tendons is dominated by diffusive
transport. In future work, we will detail models that result in precise
field values for the fluxes, which will replace the upper and lower
bounds discussed here.

This numerical example also points to the fact that a convenient
measure of the strength of 
coupling between the mechanics and mass transport equations is the
ratio of the variation in hydrostatic stress of the fluid to that of
the solid. In the lower bound case, where the fluid response is
defined by Equation~(\ref{equalpr}), it is instructive to note that
this ratio is unity. As a result, it is seen that the lower bound case
exhibits significantly weaker coupling than the upper bound case. In
the latter, variation in the common deformation gradient, $\delta
\bF$, causes instantaneous variation in \mbox{$\delta p^{\mathrm{f}} \approx
  O(\kappa^{\mathrm{f}} \delta \bF:\bF^{-\mathrm{T}})$} and in
\mbox{$\frac{1}{3} \delta\mathrm{\small{tr}}[\Bsigma^{c}] \approx
  O(\kappa^{\mathrm{c}} \delta \bF:\bF^{-\mathrm{T}})$}, where
$\kappa^{\mathrm{c}}$ is the bulk modulus of the solid. The ratio
$\frac{\delta p^{\mathrm{f}}}{\frac{1}{3} \delta
  \mathrm{\small{tr}}[\Bsigma^{c}]}$ is therefore \mbox{$\approx
O(\kappa^{\mathrm{f}}/\kappa^{\mathrm{c}}) \gg 1$}.

The strength of coupling between the equations plays a principal role
in the rate of convergence of the solution, as observed in
Table~\ref{resnorms}, where the residual norms of the equilibrium
equation (and
corresponding CPU times in seconds for an Intel\textregistered Xeon
3.4 GHz machine) are reported for the first 8 iterations of each of
the two cases. Recall that the staggered scheme involves solution of
the mechanics equation 
keeping the concentrations fixed, and the mass transport equation
keeping the displacements fixed, in turn, until the solution
converges. The table does not report the value of the residual norms
arising from the solution of the mass transport equation for the
fluid, which occurs after each reported solve of the of the mechanics
equation. Although the initial mechanics residual norms in successive
passes are decreasing linearly in both cases, the rapid decrease in
this quantity in
the weakly-coupled case ensures convergence in far fewer iterations
than the strongly coupled case. Thus, the corresponding CPU times
reported are also lower for the weakly coupled case. This is
advantageous. In addition to being more physical, as argued at
the beginning of Section \ref{swelling} immediately below, the lower
bound, weakly-coupled case makes it feasible to drive 
problems to longer, physiologically-relevant time-scales through the use
of larger time steps.

\begin{table}
\centering
\begin{tabular}{|r|c|c|c|c|}
  \hline
  Pass & \multicolumn{2}{c|}{Strongly coupled} &
         \multicolumn{2}{c|}{Weakly coupled}\\
  \cline{2-5} & Residual & CPU (s) & Residual & CPU (s)\\
  \hline\hline 
1     & $ 2.138\times 10^{-02}$ &   29.16   & $6.761 \times 10^{-04}$  &    28.5 \\
      & $ 3.093\times 10^{-04}$ &   55.85   & $1.075 \times 10^{-04}$  &    55.1 \\
      & $ 2.443\times 10^{-06}$ &   82.37   & $4.984 \times 10^{-06}$  &    81.8 \\
      & $ 2.456\times 10^{-08}$ &  109.61   & $1.698 \times 10^{-08}$  &   107.9 \\
      & $ 4.697\times 10^{-14}$ &  135.83   & $3.401 \times 10^{-13}$  &   134.1 \\
      & $ 1.750\times 10^{-16}$ &  163.18   & $1.1523\times 10^{-17}$  &   161.1 \\
\hline                                    
2     & $ 5.308\times 10^{-06}$ &  166.79   & $5.971 \times 10^{-08}$  &  192.5  \\
      & $ 4.038\times 10^{-10}$ &  193.36   & $4.285 \times 10^{-11}$  &  218.6  \\
      & $ 1.440\times 10^{-14}$ &  220.45   & $2.673 \times 10^{-15}$  &  246.1  \\
      & $ 4.221\times 10^{-17}$ &  247.04   & $                    $   &  \\
\hline                                    
3     & $ 5.186\times 10^{-06}$ &  250.62   & $2.194 \times 10^{-09}$  &  277.3  \\
      & $ 3.852\times 10^{-10}$ &  277.44   & $2.196 \times 10^{-13}$  &  304.2   \\
      & $ 1.369\times 10^{-14}$ &  304.16   & $1.096 \times 10^{-17}$  &  331.6   \\
      & $ 4.120\times 10^{-17}$ &  331.47   & $                    $   &  \\
\hline                                    
4     & $ 5.065\times 10^{-06}$ &  335.16   & $8.160 \times 10^{-11}$  &  363.2 \\ 
      & $ 3.674\times 10^{-10}$ &  362.24   & $7.923 \times 10^{-15}$  &  390.2 \\
      & $ 1.300\times 10^{-14}$ &  388.79   & $                    $   &  \\
      & $ 4.021\times 10^{-17}$ &  416.08   & $                    $   &  \\
\hline                                    
5     & $ 4.948\times 10^{-06}$ &  419.59   & $3.078 \times 10^{-12}$  &  421.4 \\
      & $ 3.503\times 10^{-10}$ &  446.24   & $3.042 \times 10^{-16}$  &  448.6 \\
      & $ 1.236\times 10^{-14}$ &  473.20   & $                    $   &  \\
      & $ 3.924\times 10^{-17}$ &  500.85   & $                    $   &  \\
\hline                                    
6     & $ 4.832\times 10^{-06}$ &  504.65   & $1.179 \times 10^{-13}$  &  479.9 \\
      & $ 3.340\times 10^{-10}$ &  531.28   & $1.291 \times 10^{-17}$  &  507.0 \\
      & $ 1.174\times 10^{-14}$ &  558.17   & $                    $   &  \\
      & $ 3.829\times 10^{-17}$ &  585.27   & $                    $   &  \\
\hline                                    
7     & $ 4.720\times 10^{-06}$ &  589.01   & $4.592 \times 10^{-15}$  &  537.8 \\
      & $ 3.184\times 10^{-10}$ &  616.24   & $5.152 \times 10^{-18}$  &  564.6 \\
      & $ 1.116\times 10^{-14}$ &  643.29   & $                    $   &  \\
      & $ 3.737\times 10^{-17}$ &  670.83   & $                    $   &  \\
\hline                                    
8     & $ 4.609\times 10^{-06}$ &  674.46   & $1.816 \times 10^{-16}$  &  595.5  \\
      & $ 3.034\times 10^{-10}$ &  701.74   & $5.040 \times 10^{-18}$  &  622.3  \\
      & $ 1.060\times 10^{-14}$ &  727.74   & $                    $   &  \\
      & $ 3.646\times 10^{-17}$ &  755.58   & $                    $   &  \\
\hline
\end{tabular}
\caption{Mechanics equation residual norms and corresponding CPU times
  in seconds for the first 8 passes of each of the two cases for a
  typical time increment, $\Delta t=$ 0.1 s.}
\label{resnorms}
\end{table}

\subsubsection{A swelling problem}
\label{swelling}

\begin{figure}[ht]
  \centering
     {\includegraphics[width=7.50cm]{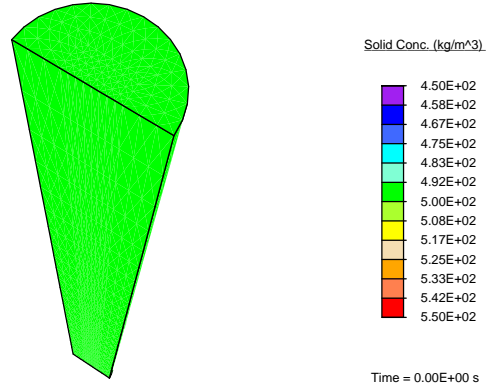}}
     \caption{The collagen initial concentration (kg.m$^{-3}$).}
     \label{before_growth}
\end{figure}

\begin{figure}[ht]
  \centering
     {\includegraphics[width=7.50cm]{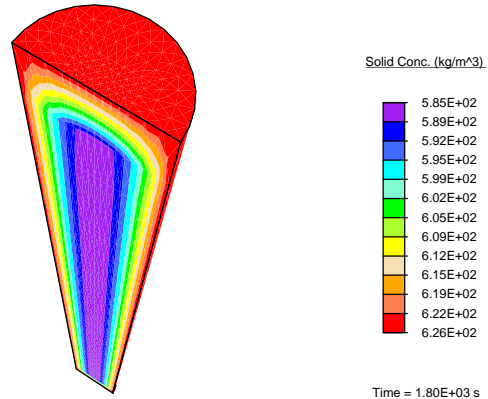}}
     \caption{The collagen concentration (kg.m$^{-3}$) after 1800~s.}
     \label{after_growth}
\end{figure}

\begin{figure}[ht]
  \centering
     {\includegraphics[angle=270,width=7.50cm]{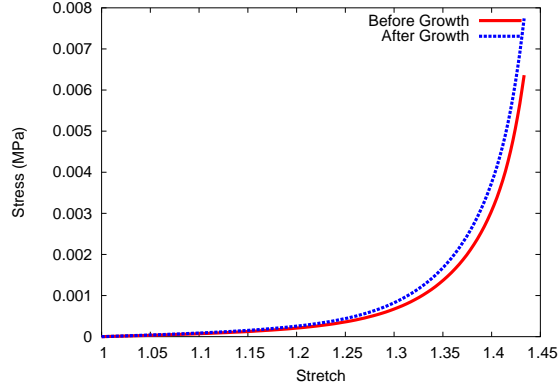}}
     \caption{The stress (Pa) vs stretch curves before and after
       growth.}
     \label{stress_strain}
\end{figure}

\begin{figure}[ht]
  \centering
     {\includegraphics[angle=270,width=7.50cm]{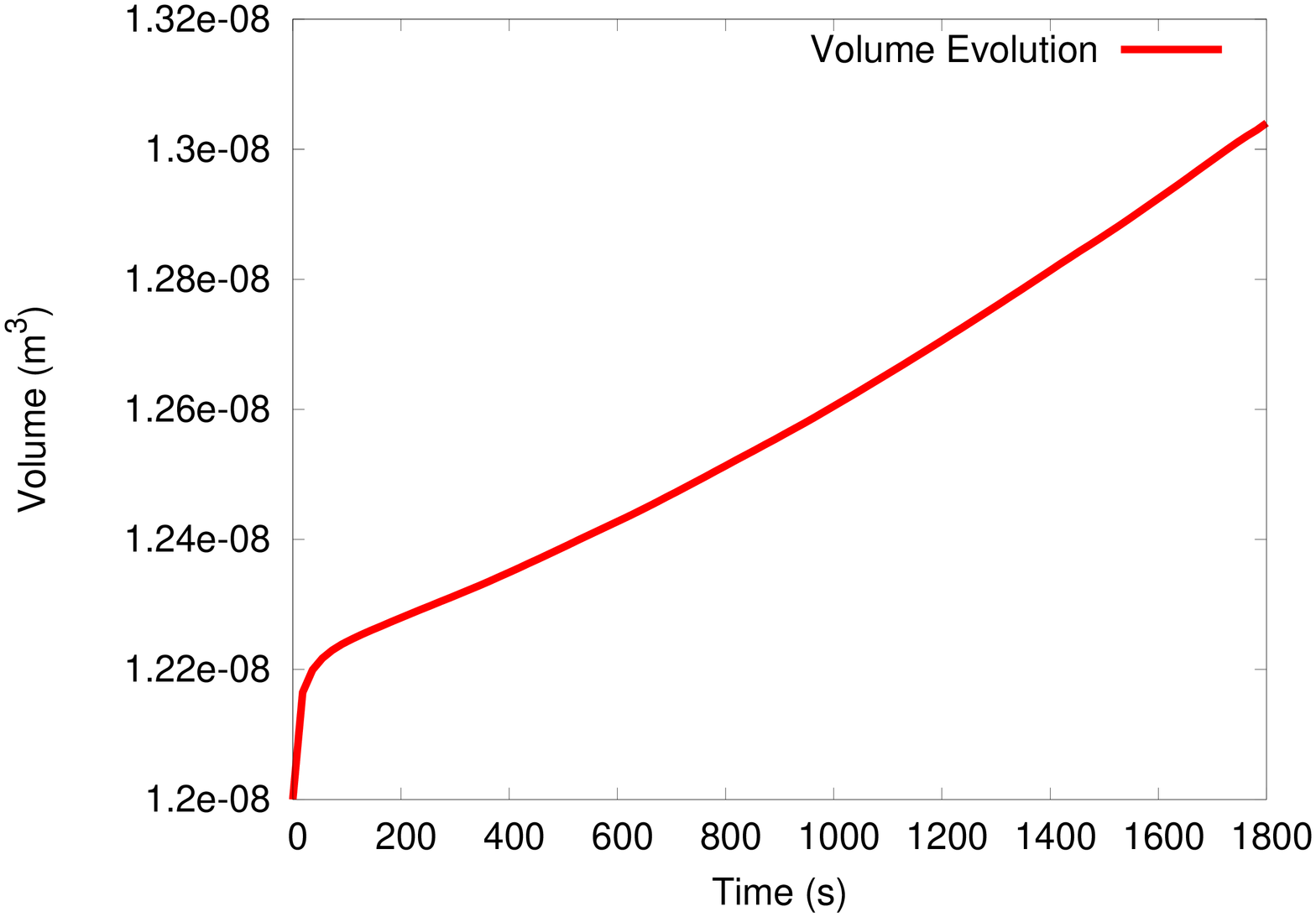}}
     \caption{The volume of the tendon (m$^3$) evolving with
     time. Note the fluid transported-dominated regime until 25 s,
     followed by the longer reaction-dominated growth stage.}
     \label{volume_evolution}
\end{figure}

Motivated mainly by the recognition that the lower bound model for
solid-fluid mechanical coupling ensures convergence to a self-consistent
solution in just a few passes of the staggered solution scheme, we
adopt this version of the coupling for our final problem. On this
  note we point out that solution of the individual balances of linear
  momentum equation for the solid collagenous and fluid phases with
  the momentum transfer terms [$\bq^\mathrm{c}, \bq^\mathrm{f}$ in
  (\ref{linearmombalance})] is a
  statement of momentum balance between them. There is reason to
  suppose, therefore, that equating the solid collagen and fluid
  stress, or some component of these tensors as done in the lower
  bound model, is a reasonable approximation to explicitly solving the
  balance of linear momentum for each phase, including the momentum
  transfers. In contrast, equating the 
  deformation gradient of the solid collagen with deformation of the
  pore spaces subjects the fluid to a stress state also determined by
  this deformation gradient in the upper bound model. This
  approximation does not correspond to an underlying physical
  principle comparable to the satisfaction of individual
  balances of linear momentum for solid collagen and fluid, with
  momentum transfers. It is therefore somewhat less motivated and more
  questionable. Clearly, a rigorous analysis or numerical
  comparisons of all three models:
  upper bound, lower bound and direct solution of individual
  solid-fluid momentum balances, must be carried out to conclusively
  demonstrate this. It is a possible topic for a future paper.

In this example we will demonstrate the mechanical
effects of growth due to collagen production. In the interest of
focusing on this issue we assume that fibroblasts are
available, and that the fluid
phase bears the necessary nutrients for
collagen production dissolved at a suitable, constant
concentration. Collagen production is assumed to be governed by a
first-order rate 
law. Newly-produced collagen has proteoglycan molecules bound to it,
and they in turn bind water. We model this effect by associating a
loss of nutrient-bearing 
free fluid with collagen production. A fluid sink $\Pi^\mathrm{f}$ is
introduced following first order kinetics,

\begin{equation}
\Pi^\mathrm{f} = -k^\mathrm{f}(\rho_0^\mathrm{f}
- \rho_{0_\mathrm{ini}}^\mathrm{f}),
\end{equation}

\noindent as in \citet{growthpaper}. Here $k^\mathrm{f}$ is the
reaction rate, taken to be 0.07 $\mathrm{s}^{-1}$, and
$\rho_{0_\mathrm{ini}}^\mathrm{f}$ is the initial 
concentration of fluid. The collagen
source is mathemaically equivalent to the fluid sink: $\Pi^\mathrm{c} =
-\Pi^\mathrm{f}$. When $\rho_{0}^\mathrm{f} >
\rho_{0_\mathrm{ini}}^\mathrm{f}$, collagen is produced.

The boundary conditions in this example correspond to immersion of the
tendon in a nutrient-rich bath. The initial collagen concentration is
500~kg.m$^{-3}$ and the fluid concentration is 400~kg.m$^{-3}$ at
every point in the tendon. When this tendon is exposed to a bath
where the fluid concentration is 410~kg.m$^{-3}$,
i.e. $\rho^\mathrm{f}(\bx,t)=410~\mathrm{kg.m}^{-3} \forall \bx \in
\partial\Omega_t$, nutrient-rich fluid is transported into the tissue,
due to the pressure difference, induced by the concentration
difference, between the fluid in the tendon and in 
the bath (fluid stress gradient-driven flux). Thereby, the nutrient
concentration is elevated, leading to collagen production, fluid
consumption and, eventually, growth due to additional collagen. 

Figure~\ref{before_growth} shows the initial collagen concentration in
the tendon. After it has been immersed in the nutrient-rich bath for
1800~s, the tendon shows growth and the collagen concentration is
higher as seen in Figure~\ref{after_growth}. On performing a
uniaxial tension test on the tendon before and after growth, it is
observed (Figure~\ref{stress_strain}) that the grown tissue is stiffer
and stronger due to its higher collagen concentration. Also note that
there is a rapid, fluid transport-dominated swelling 
of the tendon  between 0 and 25 s 
following immersion in the fluid bath
(Figure~\ref{volume_evolution}). This causes a small volume change of 
$\approx 1.6$\%. In this transport-dominated regime the contribution
to tendon growth from collagen production is small. However, the
fluid-induced swelling saturates, and between
25 and 1800 s the reaction producing collagen dominates the growth
process, producing a further $\approx 6.8$\% volume change. Noting that the
range of collagen concentration in Figure~\ref{after_growth} is
$585-626\; \mbox{kg.m}^{-3}$, and that (\ref{isotropicgrowth}) gives $\bF^{\mathrm{g}^\mathrm{c}} = \left(
  \frac{\rho_0^\mathrm{c}}{\rho_{0_{\mathrm{ini}}}^\mathrm{c}}
  \right)^  {\frac{1}{3}} {\bf 1}$, this portion of the volume change
  is quite clearly due to collagen production. The total volume change
  of $8.4$\% corresponds to changes in each linear dimension of the
  tendon by only $\approx 2.7$\%, and is not discernible upon comparing
  Figures~\ref{before_growth} and \ref{after_growth}. It is, however,
  manifest in Figure~\ref{volume_evolution}.

\section{Conclusion}
\label{sec:5}

In this paper, we have discussed a number of enhancements to our
original growth formulation presented in \citet{growthpaper}. That
formulation has served as a platform for posing a very wide range of
questions on the biophysics of growth. Some issues, such as
saturation, incompressibility of the fluid species and its influence
upon the tissue response, and the roles of biochemical and strain
energy-dependent source terms are specific to soft biological
tissues. We note, however, that other issues are also applicable to a
number of systems with a porous solid, transported fluid and reacting
solutes. Included in these are issues of current versus reference
configurations for mass transport, swelling, Fickean diffusion, fluid
response in compression and tension, cavitation and the strength of
solid-fluid coupling..

These issues have been resolved using arguments posed easily in the
framework derived in \citet{growthpaper}. The interactions engendered
in the coupled reaction-trans\-port-mechanics system are complex, as
borne out by the numerical examples in
Section~\ref{numericalimplementation}. We are currently examining
combinations of sources defined in Section~\ref{sources}, and aim to
calibrate our choices from tendon growth experiments. The treatment of
these issues has led to a formulation more suited to the biophysics of
growing soft tissue, making progress toward our broader goal of
applying it to the study of wound healing, pathological hypertrophy
and atrophy, as well as a study of drug efficacy and interaction.

\bibliographystyle{spbasic}
\bibliography{bmmb05}

\end{document}